\documentclass[iop]{emulateapj}

\usepackage{epsfig}
\usepackage{subfigure}
\usepackage{multirow} 
\usepackage{color}

\shorttitle{Modes properties in subgiants}
\shortauthors{Benomar et al.}

\begin{document}


\title{Properties of oscillation modes in subgiant stars observed by \emph{Kepler}}


\author{O. Benomar\altaffilmark{1,5}, T.R. Bedding\altaffilmark{1,5}, B. Mosser\altaffilmark{2}, D. Stello\altaffilmark{1,5}, K. Belkacem\altaffilmark{2}, R.A. Garcia\altaffilmark{3},	T.R. White\altaffilmark{1,5}, C.A Kuehn\altaffilmark{1,5}, S. Deheuvels\altaffilmark{4}, J.~Christensen-Dalsgaard\altaffilmark{5}
}

\altaffiltext{1}{Sydney Institute for Astronomy (SIfA), School of Physics, University of Sydney, NSW 2006, Australia} 
\altaffiltext{2}{LESIA, Observatoire de Paris, CNRS UMR 8109, Universit\'e Paris Diderot, 5 place J. Janssen, 92195 Meudon, France}
\altaffiltext{3}{{Laboratoire AIM, CEA/DSM-CNRS-Universit\'e Paris Diderot; IRFU/SAp, Centre de Saclay, 91191 Gif-sur-Yvette Cedex, France}}
\altaffiltext{4}{Universit\'e de Toulouse, Institut de Recherche en Astrophysique et Planétologie, UPS-OMP, CNRS, 14 avenue Edouard Belin, 31400 Toulouse, France}
\altaffiltext{5}{{Stellar Astrophysics Centre, Department of Physics and Astronomy, Aarhus University, Ny Munkegade 120, DK-8000 Aarhus C, Denmark}}

\begin{abstract}
 Mixed modes seen in evolved stars carry information on their deeper layers that can place stringent constraints on their physics and on their global properties (mass, age, etc...). In this study, we present a method to identify and measure all oscillatory mode characteristics (frequency, height, width). Analyzing four subgiants stars, we present the first measure of the effect of the degree of mixture on the $\ell=1$ mixed modes characteristics. We also show that some stars have measurable $\ell=2$ mixed modes and discuss the interest of their measure to constrain the deeper layers of stars. 
\end{abstract}


\section{Introduction}\label{sec:1:1}

Identifying oscillatory modes and measuring their properties is fundamental to asteroseismology. \\
For solar-type stars on the main-sequence, the observed oscillations are pressure waves (or p~modes), while gravity modes (or g modes) are confined to the radiative core and have a surface amplitude too small to be measured directly \citep{Belkacem2009,Appourchaux2010}. Therefore, the frequency spectrum follows the asymptotic relation for the p modes which is a rather simple pattern \citep{Vandakurov1967AZh,tassoul1980,gough1986}.

As a star evolves into a subgiant, the deeper layers of the core stop burning hydrogen which modifies significantly the internal structure and therefore the stellar oscillations. Modes that exhibit both p and g modes properties, so-called mixed modes \citep{Aizenman1977}, become visible in power spectra of such stars.

These mixed modes can be hard identify and analyze because they do not follow the asymptotic relations of pressure-modes. 
However, a recent dedicated asymptotic development has been proposed by \cite{Mosser2012a} easing the mode identification.

Mixed modes are sensitive to the deeper layers of stars and allow us to probe more precisely their evolutionary state.  For instance, rapid changes in the core of evolved stars modify the gravity-mode frequencies in very short timescales. Consequently, the frequencies of the mixed modes change quickly with time \citep{Osaki1975,
Aizenman1977}, potentially providing stellar ages with a precision down to a few Myr 
\citep{Metcalfe2010}.

Until now, the measures of the individual modes characteristics were based on assumptions valid in the Sun and similar stars (e.g. \citealt{appourchaux2008}, \citealt{Appourchaux2012}). Although these assumptions are well suited for low-mass main-sequence stars, they do not hold for more evolved stars. For example, an in-depth theoretical study of red giants by \cite{Dupret2009} showed that widths and amplitudes of the non-radial modes are smaller than for radial modes and may vary greatly from one mode to another as the modes undergo \emph{avoided crossings} (\emph{e.g.} \citealt{Deheuvels2011, Benomar2012b}). A precise measure of the properties of the non-radial modes would therefore provide unprecedented constraints on the damping and on the inertia of modes confined to the deeper layers of stars. 

The goal of this paper is to describe a global method identifying the non-radial modes and measuring all individual mode properties in subgiants observed by \emph{Kepler}, using an appropriate set of assumptions. A discussion with theoretical expectations is also presented.

We first describe the potential of mixed modes as probes of the stellar properties (Section \ref{sec:1:2}). We then develop the methods and formalism for measuring the properties of the mixed modes (Section \ref{sec:2}), define the limitations of the approach (Section \ref{sec:4:0}) and finish by presenting (Section \ref{sec:5:0}) and discussing (Section \ref{sec:6}) the results from observations.

\section{Mixed modes in evolved stars} \label{sec:1:2}

\textbf{Mixed modes have an accoustic behavior in the envelope of the star and behave as gravity waves in the deep interior; the relative importance of the acoustic and gravity-wave behavior is determined by the coupling in the mode between these two regions, which depends on the frequency and degree of the mode. }
They occur in evolved stars (subgiants and red giants), in which the large density gradient outside the core effectively divides the star into two coupled cavities. Mixed modes have p-mode character in the envelope and g-mode character in the core. This leads to {\em mode bumping}, in which mode frequencies are shifted from their regular spacing and no longer follow the asymptotic relations for p modes that is seen in main-sequence stars.

We can model mixed modes as arising from fictitious pure p and g modes that would exist if their cavities were not coupled. In order to avoid ambiguity we follow \cite{Aizenman1977} by referring to these pure modes as $\pi$ and $\gamma$ modes respectively (see also \citealt{bedding2011a}).

Mode bumping in subgiant stars was first observed and modeled in $\eta$~Boo
\citep{Kjeldsen1995b,christensen1995,Guenther1996, kjeldsen2003, carrier2005} and
$\beta$~Hyi \citep{bedding2007,Brandao2011}.  More recently, asteroseismic space missions have produced many more examples, including the CoRoT target HD~49835 \citep{Deheuvels2010a} and a growing number of \emph{Kepler} stars (e.g., \citealt{Metcalfe2010, Mathur2010, campante2011, Benomar2012a, Appourchaux2012}). Meanwhile, thousands of giants have been observed with CoRoT and \emph{Kepler} showing a `forest' of non-radial modes \citep{Bedding2010, Huber2010, Kallinger2010} 
identified as $\ell=1$ mixed modes. This has led to two major discoveries: a seismic diagnostic that permits us to distinguish RGB stars from red clump stars \citep{Beck2011Science, Bedding2011Nature, Mosser2011a} and evidence for radial differential rotation within stars \citep{Beck2012Nature, Deheuvels2012,Mosser2012c, Marques2013, Goupil2013a}.

Although all non-radial modes can have mixed properties, most work to date on subgiants and giants has focused on dipole modes ($\ell=1$) because the distance between the pressure and gravity wave cavities is less than for higher-degree modes. \textbf{Hence the trapping is less efficient for dipole modes than at higher degrees. Observationally it means that dipole modes lifetimes are smaller and their amplitudes higher \citep{Dupret2009}; therefore they are more easy to resolve and to observe.} 

In the early subgiant phase, only few avoided crossings are observed because the $\gamma$ modes are well spaced in frequency \textbf{(low $\gamma$-mode density)}. As a subgiant evolves, the $\gamma$-mode \textbf{density increase}, thus the number of $\gamma$ modes within the observable range of frequency increases 
\textbf{(approximately between $0.5\,\nu_\mathrm{max}$ and $1.5\,\nu_\mathrm{max}$, where $\nu_\mathrm{max}$ is the frequency at maximum power for the modes)}. This leads to numerous (measurable) avoided crossings. With more avoided crossings, the spectrum becomes very complicated, which makes mode identification difficult, because mixed modes spread all over the \'echelle diagrams. When the number of $\pi$ and $\gamma$ modes become similar, the spectrum is particularly difficult to interpret.

Recently, theoretical developments \cite[in prep.]{Goupil2013} following the formalism of \cite{Unno1989}, have led to an asymptotic relation for mixed modes, which has been applied by \cite{Mosser2012a} to giant stars. With this asymptotic expansion, modes can be more easily identified in giants and subgiants. The asymptotic relation for mixed modes uses constant values of the large separation ($\Delta\nu$) and of the period spacing ($\Delta{\Pi_\ell}$), to characterize the mixed modes and is well suited to measuring global frequency  characteristics of the mode bumping. However it does not provide other mode properties (e.g. height or width of the modes). 

The approach presented in this paper aims to fit all individual properties of the modes. It describes the frequencies of $\ell=1$ modes as eigensolutions of a series of coupled harmonic oscillators, which has been shown to fit observed avoided crossings \citep{Deheuvels2011, Benomar2012a}. The approach uses a full Bayesian framework to include our prior knowledge of the pressure and gravity modes. 

 \section{Fitting the power spectrum} \label{sec:2}
 
 In this section, we present the model that describes the frequencies, referred as the coupled-oscillators model and the method and assumptions for the fit of the spectra. 
 
 \subsection{The Coupled-Oscillators Model} \label{sec:2:1}

\textbf{In the coupled oscillator model we expand the oscillations of a stellar model on fictitious pure $\pi$ and $\gamma$ modes, with frequencies $\omega_j^{(\pi)}$, $j=\left\{1, \ldots, N_{\pi} \right\}$, and $\omega_k^{(\gamma)}$, $k=\left\{1, \ldots, N_\gamma\right\}$, respectively. We seek angular frequencies $\omega=\{\omega_1, \ldots, \omega_N\}$, with $N=N_\pi + N_\gamma$, which are to be fitted to the observed frequencies. This leads to the following system of equation,}
\begin{equation} \label{eq:LS}
\mbox{\boldmath$A$} Y=\omega^2 Y,
\end{equation} 
where $Y=\left\{ y_1, y_2, ..., y_N \right\}$ is a vector \textbf{of mode amplitudes} and $\mbox{\boldmath$A$}$ is a matrix that contains the frequencies of the $\pi$ and $\gamma$ modes, as well as the coupling terms $\alpha$
\begin{equation} \label{eq:mat_system}
\mbox{\boldmath$A$}= \left( \begin{array}{ccccccc}
({\omega^{(\pi)}_1})^2 & \cdots     &    0                      &    -\alpha_1           &  \cdots     & -\alpha_{N_\gamma}      \\
     \vdots      & \ddots     &    0                      &    -\alpha_1		 	 	   &  \cdots     & -\alpha_{N_\gamma}         \\
    0            &  \cdots    & ({\omega^{(\pi)}_{N_\pi}})^2     &    -\alpha_1           &  \cdots     & -\alpha_{N_\gamma}      \\
  -\alpha_1      &  \cdots    & -\alpha_1                 &    ({\omega^{(\gamma)}_{1}})^2 &  \cdots     &        0          \\
  \vdots         &  \cdots    & \vdots                    &    \vdots &  \ddots     &        0          \\
  -\alpha_{N_\gamma}      &  \cdots    & -\alpha_{N_\gamma}           &       0                &  \cdots     & ({\omega^{(\gamma)}_{N_\gamma}})^2 
\end{array} \right).
\end{equation}
\textbf{Note that the coupled oscillator does not represent a physical model of stellar oscillations but it does contain the salient aspects of the properties of the mixed modes, allowing it to provide a description of the observed frequencies in terms of the relevant properties of the star. }
Following \cite{Deheuvels2010b} and \cite{Benomar2012b}, we assumed for each $\gamma$ mode (that is, at each avoided crossing), that the coupling strength between this $\gamma$ mode and the $N_\pi$ $\pi$ modes has a single value, $\alpha_i$, with $i=\left\{1, ..., N_\gamma \right\}$.
Note that at early stages of the sub-giant phase, $N_\pi \gg N_\gamma$: the spectrum contains few avoided crossings and has a p-mode dominated pattern \citep{Deheuvels2011, campante2011, Appourchaux2012}. At the other extreme, red giants have $N_\pi \ll N_\gamma$ and the power spectrum has a g-mode dominated pattern \citep{Dupret2009, bedding2011a, Stello2011, christensen2011b, Mosser2012a}. Between these two
extremes, one can consider intermediate cases where $N_\pi \approx N_\gamma$ \citep[e.g.,][]{DiMauro2011}, whose power spectra may be hardest to interpret.

	\subsection{Markov Chain Monte Carlo fitting} \label{sec:2:2}
	
	In recent years, Markov Chain Monte Carlo (MCMC) approaches have been widely used in asteroseismology (e.g. \citealt{Brewer2007, Gruberbauer2009, benomar09, benomar09b, Handberg2011, campante2011, Appourchaux2012}). Such approaches rely on sampling the posterior Probability Density Function (PDF) and provide far more robust results than maximization methods such as the Maximum Likelihood Estimator (MLE) or the Maximum A Posteriori (MAP) approach. 
	
MCMC fitting is well suited to the present analysis because it provides an easy way to compute the PDF of any function 
of the fitted variables. 
For example, the PDF of the mean large separation or the mean period spacing (cf. Fig. \ref{fig:Simu:correlation}) can be easily derived from the samples acquired by MCMC, for each individual p-mode or g-mode frequencies. 
When dealing with very complex functions or highly non-linear models, such as the coupled oscillators model, it provides  an easy way to propagate the uncertainties.
Note that the matrix $A$ contains often hundreds of correlated parameters, so the error propagation law would require us to determine thousands of derivative terms, either analytically or numerically. MCMCs allow one to propagate the error in a more straightforward and simpler way. Moreover it does not rely on the PDFs being Gaussian.
	
	\subsection{Defining the posterior probability density function} \label{sec:2:3}
	
	In a Bayesian approach, the first step is to write the likelihood function and define the posterior PDF. We deal with a power spectrum, thus the noise statistic of each data point $y_i$ follows a $\chi^2$ with two degrees of freedom. Knowing the statistics of data points is sufficient to derive the likelihood function, given the fitted model $M(\nu, \mbox{\boldmath$\theta$})$ that depends on variables \mbox{\boldmath$\theta$} \citep{duvall1986}:
\begin{equation}\label{eq:likelihood}
 p(y|\mbox{\boldmath$\theta$},M,I)=\prod_{i=1}^N \frac{1}{M(\nu_i,\mbox{\boldmath$\theta$})}\mathrm{e}^{-\frac{y_i}{M(\nu_i,\theta)}},
\end{equation}
where $N$ is the number of frequencies $\nu_i$ and $I$ encompasses all other contextual information. 
Equation (\ref{eq:likelihood}) is only applicable if the sampled frequencies are independent, that is, if the power spectrum is sampled at the formal frequency resolution of the data set.

From Bayes theorem, we can define the posterior PDF that we seek to sample,
\begin{equation}
 p(\mbox{\boldmath$\theta$}|y,M,I) = p(\mbox{\boldmath$\theta$}|M,I) p(y|\mbox{\boldmath$\theta$},M,I)/ C 
\end{equation}
where $p(\mbox{\boldmath$\theta$}|M,I)$ is our explicit prior knowledge on the parameter set \mbox{\boldmath$\theta$} of the model $M$, and $C$ is a normalization constant.
To apply priors on a particular parameter (or a subset of parameters), we use the product rule for the case of independent variables: the global prior is simply the product of the individual priors. For example, we can separate the prior associated with the noise \mbox{\boldmath$\theta_N$} from the prior on the mode \mbox{\boldmath$\theta_S$},
\begin{equation} p(\mbox{\boldmath$\theta_S$},\mbox{\boldmath$\theta_N$}|M,I)=p(\mbox{\boldmath$\theta_S$}|M,I) p(\mbox{\boldmath$\theta_N$}|M,I).
\end{equation}

	\subsection{Defining the priors} \label{sec:priors}

		\subsubsection{Frequencies} \label{sec:priors:freq}
		
To a first approximation, each resolved mode is well represented by a Lorentzian in the power spectrum. This Lorentzian is characterized by a height $H$, a width $\Gamma$ and a central frequency $\nu$. 
\textbf{The frequency distribution of $\pi$ modes is approximately described by the second-order asymptotic relation for pressure modes \citep{tassoul1980},}
\begin{equation} \label{eq:ass_law}
	\nu_\pi(n_\pi) \simeq \left( n_\pi + \frac{\ell}{2} + \epsilon_\pi \right) \Delta\nu - \delta\nu_{0\ell}.
\end{equation}
\textbf{where $\delta\nu_{0\ell}$ represents the second-order term and $\epsilon_\pi \simeq 1/4$ for low-mass stars, as verified by \cite{Mosser2013a}. For an accurate description, one must take care to distinguish the observed and asymptotic values of the large separation.}
For $\gamma$ modes, the asymptotic relation for periods $P_\gamma$ is \citep{tassoul1980} 
\begin{equation} \label{eq:ass_law2:1}
	P_\gamma(n_\gamma, \ell) = \left\{ 
		\begin{array}{ll}
				\left( n_\gamma + \frac{\ell}{2} + \alpha_{\gamma} \right) \Delta{\Pi_\ell}, &\mbox{ (radiative core)} \\
			  \left( n_\gamma + \alpha_{\gamma} \right) \Delta{\Pi_\ell}, &\mbox{ (convective core).}
		\end{array}
		\right.
\end{equation} %
Here $n_\pi$ and $n_\gamma$ are the radial orders for $\pi$ modes and $\gamma$ modes, $\ell$ is the degree, $\epsilon_\pi$ is a phase offset related to the position of the waves turning point in the stellar atmosphere (\emph{e.g.}~\citealt{White2011}), while $\alpha_{\gamma}$ and $\delta\nu_{0\ell}$ carry information about the core. $\Delta{\Pi_\ell}$ is the period spacing of $\gamma$ modes with degree $\ell$ such as $\Delta{\Pi_\ell} = \Delta{\Pi_0} / \sqrt{\ell (\ell + 1)}$. 

These equations are used in order to define the priors on the frequencies of the $\pi$ and $\gamma$ modes. Note that, in practice, $\alpha_{\gamma}$ cannot be directly measured as we can't distinguish \emph{a priori} between convective and radiative cores. Therefore, we will use $\epsilon_{\gamma,\ell}$ to denote the phase offset for the gravity-modes, independent of the nature of the core. 

 Abrupt changes in the physical properties inside stars, known as glitches, cause departures from the predicted asymptotic frequencies for both pressure and gravity modes. We expect the departures to be less than a few percent of $\Delta\nu$ \citep{Provost1993, Mazumdar2010, Miglio2010}.
 
In order to take these glitches into account (and to reproduce accurately the observed mixed mode frequencies), we used an approach  similar to \cite{Benomar2012a} for the $\pi$ modes. We applied smoothness conditions on the average large separation, as well as on the second derivative in $n_\pi$ for the frequencies $\nu_\pi(n_\pi)$. The priors on $\Delta\nu$ ensure that the $\ell=1$ $\pi$ modes stay approximately aligned in the \'echelle diagram, while the prior on the second derivative avoids strong, non-physical kinks.  This was achieved by applying a Gaussian prior on these quantities. For example, for the second derivative in frequency $\Delta^2\nu (n)$, we have,
\begin{equation}
	p( \Delta^2\nu (n) | M, I) = \frac{1}{\sqrt{2 \pi} \sigma_{\Delta^2\nu}} \exp\left[- \displaystyle{1\over 2} \left(\frac{\Delta^2\nu (n)}{\sigma_{\Delta^2\nu}} \right)^2 \right],
\end{equation}

where $\sigma_{\Delta^2\nu}$ plays the role of a relaxation constraint and must
be chosen to ensure enough freedom, but not too much, in order to
efficiently smooth the frequency profile.  Here, a trial-and-error procedure
showed that values of $\sigma_{\Delta^2\nu}$ between $0.5$ to $1.5$ $\mu$Hz offer a good compromise: stars with a greater number of mixed modes need a tighter prior.

In a very similar way, smoothness conditions were applied to the $\gamma$ modes, with the difference that they were applied to the periods of the modes (period spacings and second derivative of the periods), instead of frequencies. In subgiants the number of $\gamma$ modes within the range of observed frequencies is small (\emph{i.e.} $N_\pi \gtrsim N_\gamma$). Therefore to avoid overfitting, the relaxation constraint was set such that mode-to-mode variations of more than $1\%$ in period spacing were unlikely.

With such smoothness conditions, the $\pi$-mode and $\gamma$-mode deviation from a strictly regular pattern of frequencies or periods is
locally described by a second-order polynomial function of the radial order, and the solution belongs to the family of spline functions. 
	
 \subsubsection{Other parameters} \label{sec:priors:others}
 
We also need to define priors for the mode linewidths. Mode bumping not only affects modal frequencies, but also their widths. Indeed, the higher the mode amplitude in the $\gamma$-cavity, the smaller the mode linewidth \citep[e.g.][]{Dupret2009}. Moreover,  mixed modes close to the $\ell=1$ $\pi$-mode \emph{`home ridge'} (\emph{i.e.} the expected frequencies of pure p modes), have more pressure-like properties than more strongly bumped mixed modes, which have more gravity-like properties. Thus, lifetimes of mixed modes are expected to vary significantly, and will not follow the usual profile seen for pure p modes in main-sequence stars (a decreasing function of frequency, with a central plateau near $\nu_\mathrm{max}$). 

For dipolar modes ($\ell=1$), it is reasonable to use the adjacent $\ell=0$ widths as an upper limit because the mixed modes should be narrower than pure p modes. Here, we used a smooth edge (rather than an abrupt one, as it would be with a uniform prior), 
\begin{equation} \label{eq:ass_law2:2}
	P(\Gamma_1) = \left\{ 
		\begin{array}{ll}
				0 \mbox{   if } \Gamma_1 < 0, \\
			  C \mbox{   if } 0 \le \Gamma_1 \le \Gamma_0, \\
			  \exp^{-\frac{1}{2}(\Gamma_1 - \Gamma_0)^2 / \sigma^2} \mbox{ if } \Gamma_1 > \Gamma_0, 
		\end{array}
		\right.
\end{equation} %
where $C=(\Gamma_0 - \Gamma_1 + \sqrt{2\pi}\sigma/2)^{-1}$, ensures a proper normalization and $\sigma=0.1 \Gamma_0$

 For $\ell \ge 2$ modes, we assumed that the linewidths follow those of the $\ell=0$ modes and we interpolated them.
This is justified by  the fact that pressure and gravity modes for $\ell=2$ modes are less coupled because the width of evanescent region between the cavities is larger than for $\ell=1$ modes \citep[e.g., Eq. (16.51) of][]{Unno1989}. 
Moreover, in the $\gamma$ cavity the radiative damping increases as $\ell^2$ \citep[e.g.,][]{Godart2009}. Mixed-mode amplitudes decrease quickly as $\ell$ increases and modes far from their home ridge are difficult to observe for $\ell \ge 2$. Consequently, for $\ell \ge 2$, observed modes are likely to be p-dominated and exhibit mode lifetimes comparable to $\ell=0$ modes \citep{Dupret2009}. 

For mode heights, which are correlated with their widths, we chose to simply use Jeffrey (non-informative) priors (it would introduce biases to use informative priors for both widths and heights).

The noise background characterization can be critical when dealing with low signal-to-noise data. Convective motions at the surface of a star are in general the main sources of noise when we analyze the modes of solar-like stars in photometry. In the present study, two Harvey-like profiles \citep{harvey1985} plus white noise are used in order to account for the different scales of the convection cells (\emph{e.g.} \citealt{Mathur2011}). Priors on the noise parameters were based on a fit where the mode envelope was described by a Gaussian function, \emph{e.g.} as in \cite{benomar09}.

Rotation and magnetic fields are known to lift the degeneracy on the degree $\ell$ such that each degree shows a fine structure and is split into $2\ell +1$ modes. In most cases the magnetic field is negligible, so this splitting is interpreted as a signature of stellar rotation (\emph{e.g} \citealt{Gizon2003, Deheuvels2012,Mosser2012c, Marques2013, Goupil2013a}). Even if the rotational splitting could be measured, we preferred not to implement it in this analysis. Indeed, several effects may make the rotational splitting hard to measure. First of all, it is strongly correlated with the inclination angle between the line of sight of the observer and the rotation axis of the star \citep{Gizon2003}. Moreover, as for the width, it varies as function of the frequency for mixed modes \citep{Beck2012Nature, Deheuvels2012}. Finally, the height of the split components could be affected by the mixed properties of the modes. Incorporating all these possible effects would greatly complicate the fitted model and goes beyond the scope of this work. 

\subsection{Defining the initial guesses} \label{sec:method}

Before fitting of the power spectrum, one needs to define the initial guesses of the fitted model. Because models often have hundreds of parameters, several steps are necessary:
\begin{itemize}
	\item[1)] We used the algorithm described by \cite{Benomar2012a} to measure the noise characteristics, the mode envelope parameters, the mean large separation $\Delta\nu$, as well as $\epsilon_p$, the phase offset for $\ell=0$ p~modes.
	\item[2)] After normalizing by the noise background the smoothed power spectrum, we identify peaks of power for which a null hypothesis is rejected at $5\%$ and consider them as potential modes \citep{Appourchaux2003}. 
	\item[3)] We assigned an identification of $\ell$ and $n$ for each significant peak that was compatible with Eq.\ref{eq:ass_law} (p modes) or that follows the expected pattern of mixed modes using the asymptotic relation for the mixed modes \citep{Mosser2012a}. This gave us a first guess for $\Delta\Pi_1$ and we checked the rendering in a replicated \'echelle diagram, as in \cite{Benomar2012b}.
	\item[4)] A pre-fit (using a MAP approach) to the $\ell=1$ peaks provided initial guesses for the matrix \mbox{\boldmath$A$} (Eq.\ref{eq:mat_system}). Several iterations were sometimes needed between steps 2 and 3, before getting a satisfactory result. Particular attention was given to the number of $\pi$ and $\gamma$ modes, in order to accurately reproduce the avoided crossings. 
	\item[5)] Guesses on widths of the modes ($\Gamma$) were provided by the correlation function between $\epsilon_p$ and $\Gamma$, found by \cite{White2012}. Initial heights were defined with the help of the envelope height and $\Delta\nu$, as demonstrated by \cite{Benomar2012a}.
	\item[6)] The final step was to define priors on the observed frequencies. In this paper, we used uniform priors, set after a careful examination of the power spectrum. These priors avoided mode mis-identification (in radial order and degree), and sped up the sampling.
\end{itemize}

Steps (3), (4) and (6) rely partly on a visual inspection of the power spectra. This inspection is important when the spectra show many mixed modes (e.g. giant stars) and/or in low signal-to-noise modes. We note however, that the asymptotic relations for the p modes and for the mixed modes have strong theoretical ground, and therefore give confidence to our identifications. Moreover, simulations (Section \ref{sec:4:0}) strengthen the validity of our method.

\section{Tests on simulated spectra} \label{sec:4:0}

	\textbf{The method presented above allows us to fit the properties of the modes that are statistically significant in the power spectrum, namely,}
	\begin{itemize}
		\item[-] \textbf{The frequencies, heights and widths of the observed modes of degree $\ell={0,1,2,3}$.}
		\item[-] \textbf{The frequencies of the fictitious $\ell=1$ $\gamma$ and $\pi$ modes as well as their coupling coefficient $\alpha$.}
	\end{itemize}
	
	In order to verify the robustness of our method, we applied it to spectra generated artificially that mimic the main characteristics (modes height, width, frequencies as well as the signal-to-noise) of real stellar cases.
		\subsection{Frequencies}
	
	We found the observed frequencies (directly observed p modes and mixed modes), and the input frequencies to agree within $2\sigma$ with no evidence of biases. 
\textbf{It is interesting to compare the mean large spacing $\Delta\nu$ and the mean period spacing $\Delta\Pi_1$ with the input values of the simulated spectra because they are good indicators of the mass, radius and age of the stars.
	Concerning the inferred $\ell=1$ $\pi$-mode frequencies, the approach works well when the density of $\gamma$ modes is lower than the density of $\pi$ modes. For the less complex cases (when the number of $\gamma$ modes per $\Delta\nu$ was less than $\approx 1$ at $\nu_\mathrm{max}$), the method accurately extracts the departures that arise from glitches in the stellar structure. }	
\textbf{However, when the density of the $\gamma$ modes was higher than that of the $\pi$ modes (\emph{i.e.} during the early giant phase, when $\Delta\nu \approx 20 - 35$ $\mu$Hz) the amount of information that comes from the power spectrum was not high enough to ensure the accuracy.}

\begin{figure*}
\includegraphics*[angle=90,totalheight=6.8cm]{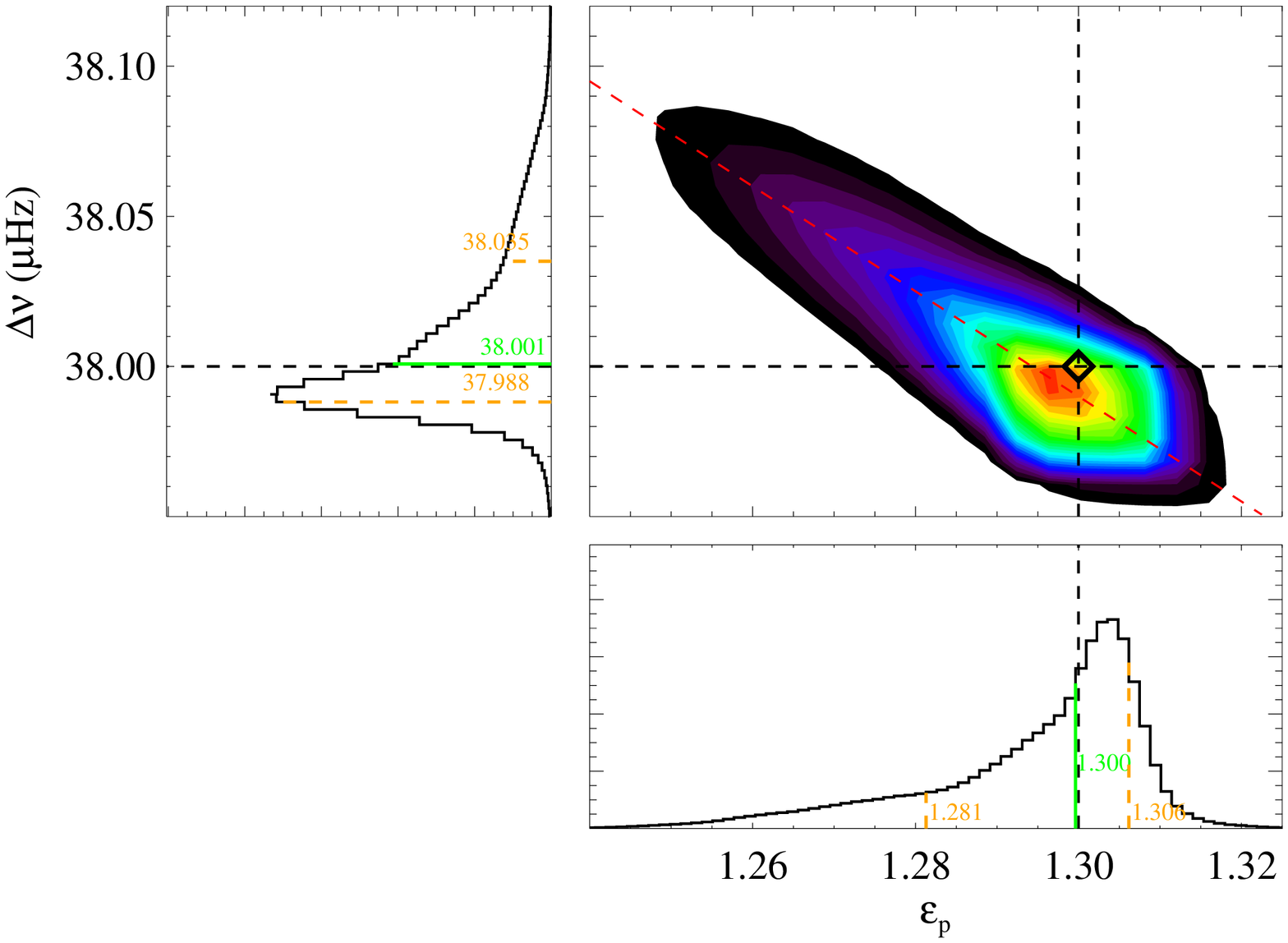} 
\includegraphics*[angle=90,height=6.8cm]{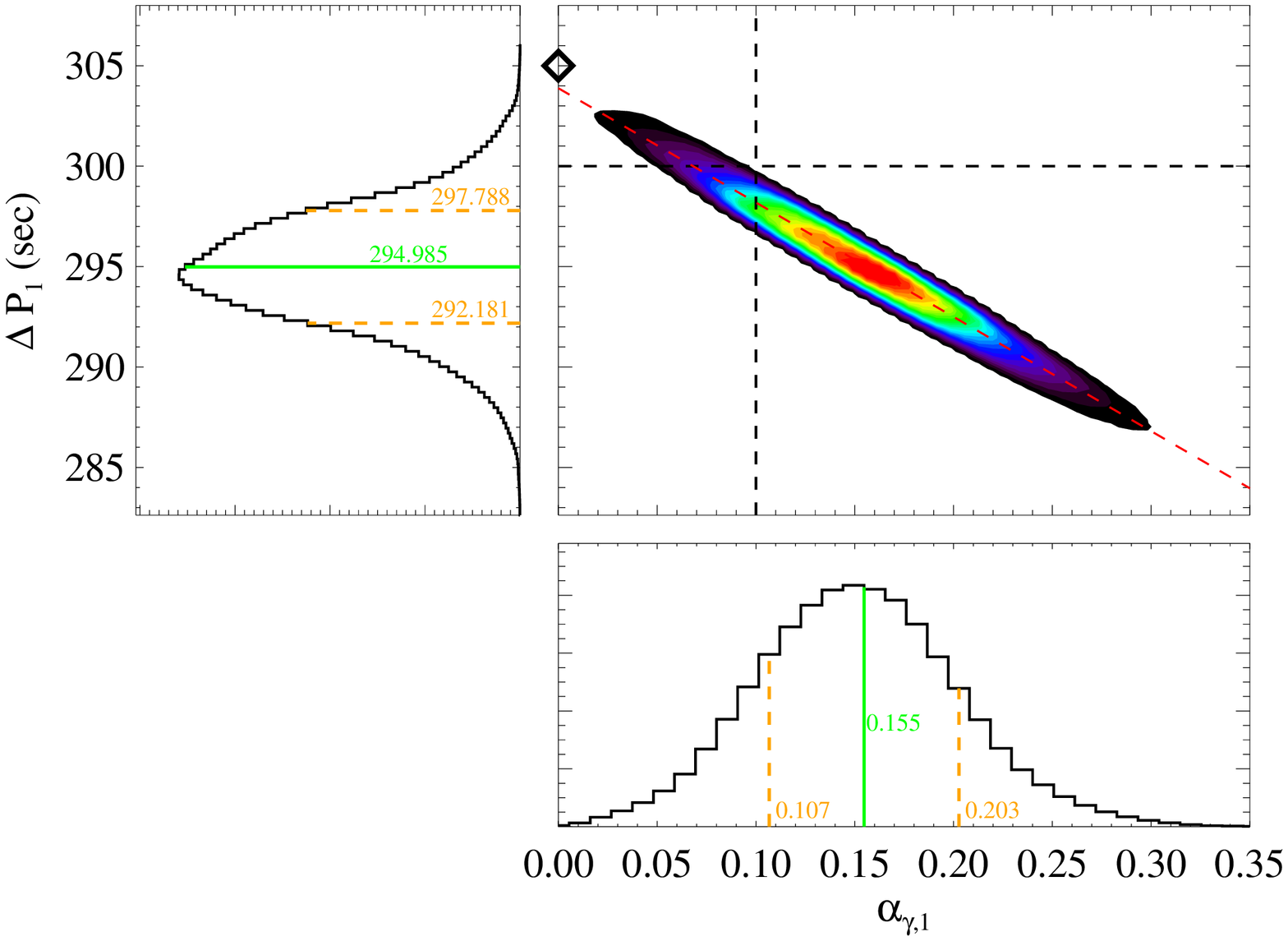}  \\
\caption{Correlation function and the PDFs for the large separation and $\epsilon_p$ (left), for $\Delta{\Pi_1}$ and $\alpha_{\gamma,1}$ (right) measured with the coupled oscillators model on a synthetic spectra. Black dashed lines indicate the true values of the parameters. The red dashed line is a linear fit of the correlation function. The value retrieved with the asymptotic relation for the mixed modes is shown with a black diamond. The correlation introduces quite often biases of a few percent. This bias can be as large as $7\%$ for $\Delta{\Pi_1}$. Fixing $\alpha_{\gamma}$ or $\epsilon_p$ to their true value removes bias. }
\label{fig:Simu:correlation}
\end{figure*}

Fig. \ref{fig:Simu:correlation} shows the correlation function and the PDFs for $\Delta\nu$ and $\epsilon_p$, and for $\Delta{\Pi_1}$ and $\alpha_{\gamma}$. This illustrates the degree of precision achievable for these global quantities.
It is clear that the correlations reduce the precision on these parameters, and may introduce systematics. The observed difference between the true value and the measured one is relatively small for $\epsilon_p$ and $\Delta\nu$ (within $1\sigma$ confidence interval), because p modes are directly observed in the power spectrum. 
Concerning the $\ell=1$ $\gamma$ modes, our lack of knowledge about $\alpha_\gamma$ and the difficulty to infer it (and therefore $\Delta{\Pi_1}$) means that the uncertainties on $\Delta{\Pi_1}$ and $\alpha_\gamma$ are of about $1$ to $5 \%$. In addition, we notice slight systematic error, that remains less than $4\%$. The eigensolutions of $\mbox{\boldmath$A$}$ (Eq.\ref{eq:LS}) change slightly as a function of the number of implemented modes in the simulation. \textbf{Neglecting the lowest frequency $\gamma$ modes shifts eigenfrequencies towards lower values and overestimates $\alpha_\gamma$ and underestimates $\Delta{\Pi_1}$}. The effect is intrinsic to the adopted methodology and to the coupled-oscillators model approach. Ideally, we should use the full set of $\gamma$ modes, which follow the asymptotic relation of the g modes, to ensure a very accurate fit. However in practice this is impossible, due to computational limitations. In order to limit this bias as much as possible, we paid careful attention to this issue by checking when eigensolutions are virtually unchanged by adding/substracting $\pi$ or $\gamma$ modes (see Section \ref{sec:method}). 
Finally, we also compared our $\Delta\Pi_1$ and $\Delta\nu$ values to those obtained with asymptotic relation for mixed modes (cf. \citep{Mosser2012a}) and found that our two approaches are in agreement (cf. Fig. \ref{fig:Simu:correlation}).

\textbf{With all the assumptions presented above, the coupled oscillator model provide a good guide to identify and fit the mode frequencies in evolved solar-like stars.}

			\subsection{Height, width and their dependence on the splitting and inclination} \label{sec:4:1}
			
The conditions for which the fit is reliable depend on whether the assumptions of Section \ref{sec:priors} are fulfilled. The most important assumption concerns the rotational splitting.
Our simulations show that the splitting could not be measured in two situations:
	\begin{itemize}
			\item[1)] When the splitting was smaller than the width of the modes and could not be resolved. 
			\item[2)] When the inclination angle $i$ was below about $30^\circ$, so that the $m \neq 0$ components contained only a small fraction ($\lesssim 10 \%$) of the total power of the modes and each degree $\ell$ was well approximated by a single Lorentzian. 
	\end{itemize}

\begin{figure*}
\includegraphics*[angle=90,totalheight=4.4cm]{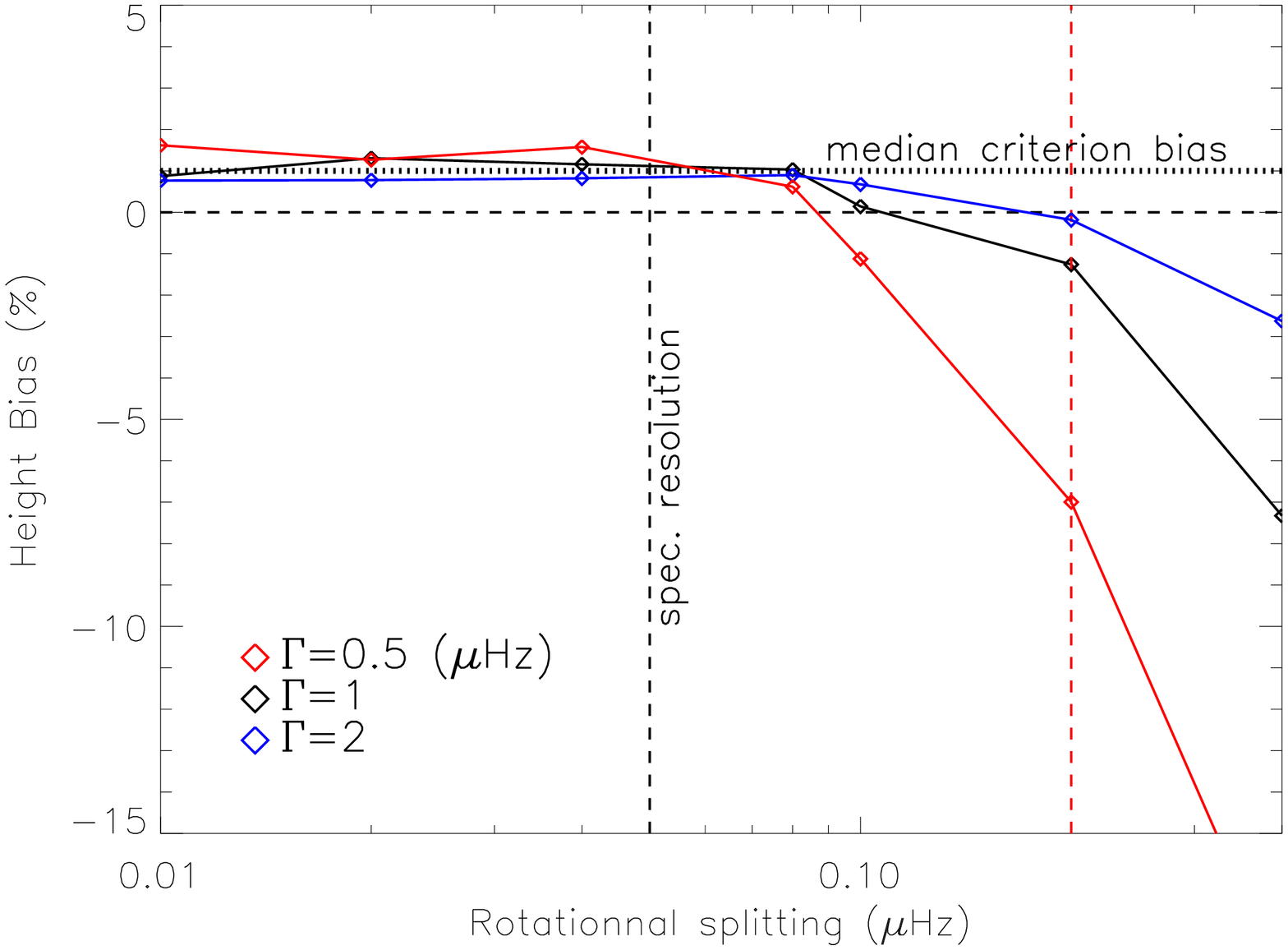} 
\includegraphics*[angle=90,totalheight=4.4cm]{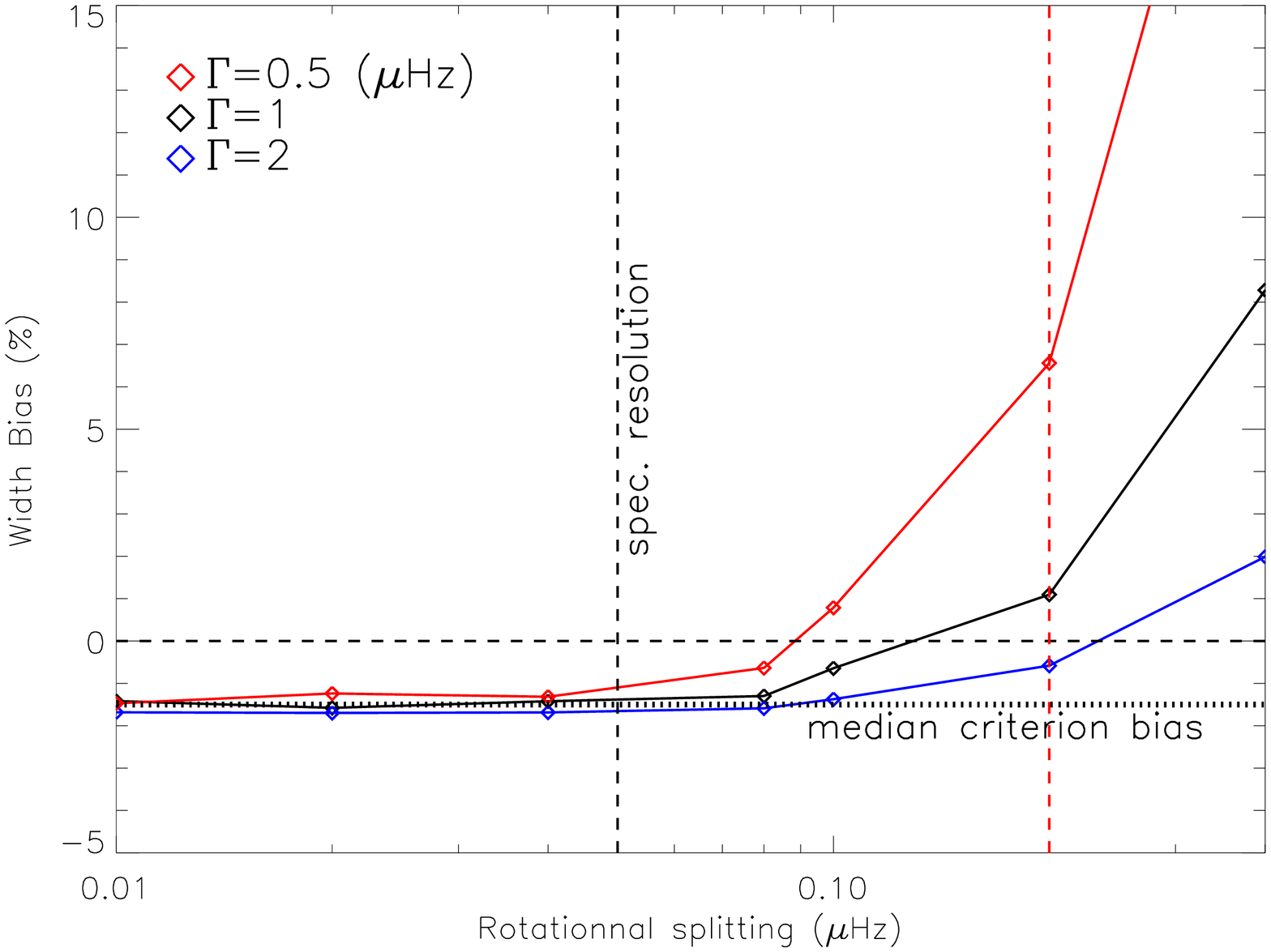}
\includegraphics*[angle=90,totalheight=4.4cm]{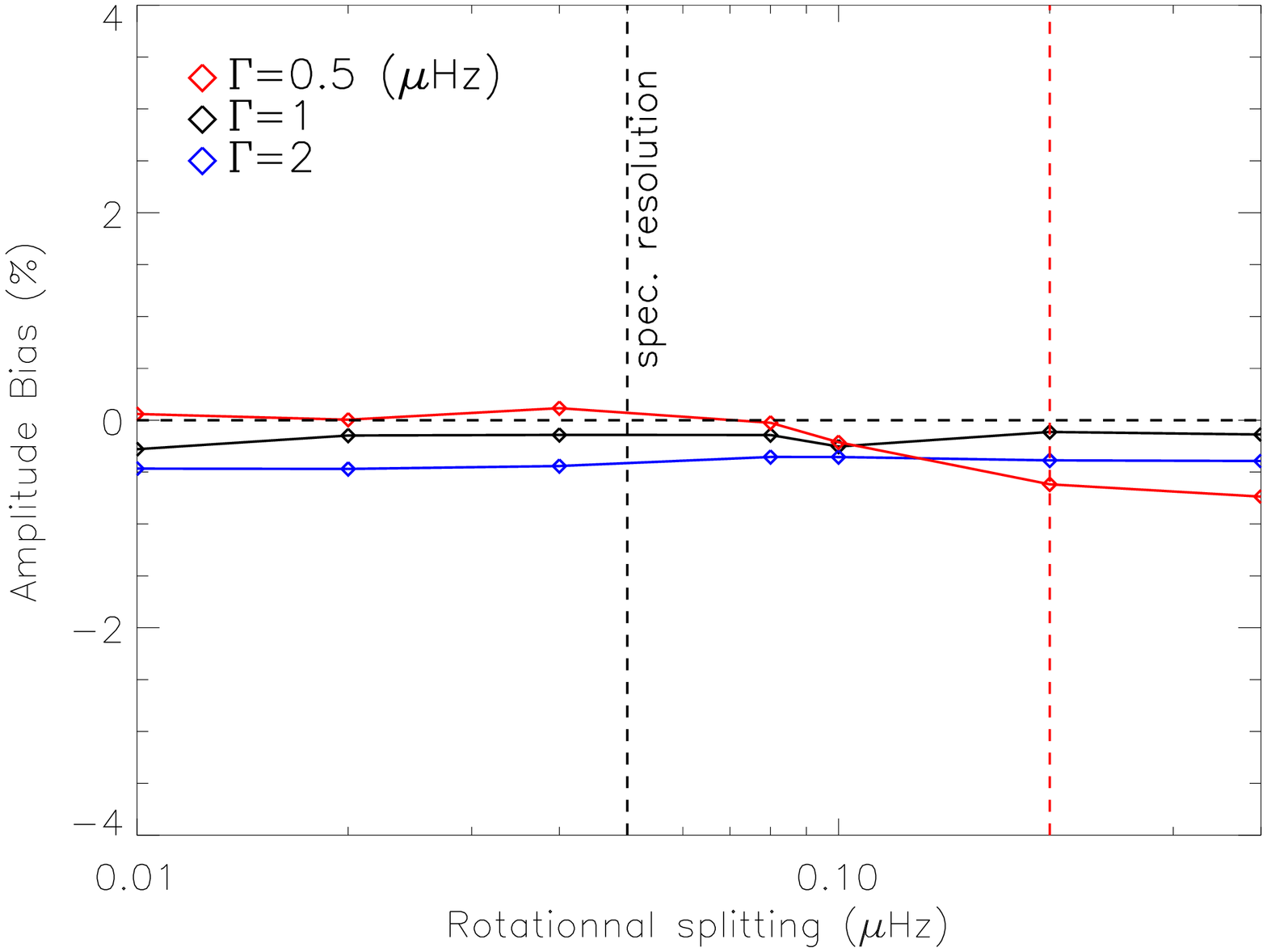}    \\
\caption{\textbf{Bias on height, width (FWHM, $\Gamma$) and amplitude while fitting a $\ell=1$ triplet (inclination of $90^\circ$) with a single Lorentzian as a function of the rotational splitting. The red dotted vertical line indicates the lower limit to visually observe the splitting in the power spectrum for $\Gamma=1$ $(\mu$Hz). The bias is more important on height and width for narrow modes. Amplitudes are almost not biased. There is a residual bias (ie. when the splitting is of $0$ $\mu$Hz) due to used statistical criterion (median).}}
\label{fig:Simu:Biases}
\end{figure*}
			When one of these conditions was fulfilled, the measured height and width of a given degree \textbf{has a minimal bias}.
			
			\textbf{To evaluate the bias due to neglecting the rotational splitting, we fitted a limit-spectrum (without noise) of an $\ell=1$ triplet at an inclination of  $90^\circ$ with a single Lorentzian. This is a worst-case scenario because the energy is only on the $m=\pm 1$ components. Fig. \ref{fig:Simu:Biases} shows the results of such a fit.  The MCMC analysis provides PDF of the fitted parameters, and a criterion has to be chosen as `best' fit. Like in \cite{benomar09, benomar09b} and \cite{Gruberbauer2009}, we chose the median because it is more robust than the mean or the maximum of the PDF for noisy data. This statistical criterion induces a bias of about $1\, \%$ the height and the width\footnote{\textbf{The maximum of the PDF is a better estimator of the height and the width while fitting a limit-spectrum, but is not suitable with noisy data because the PDF may have several maxima. We also remind that Bayesian approaches coupled with MCMC are global optimisation approaches and one should rely on the confidence intervals rather than on an arbitrary criterion such as the median.} }. Knowing this, the bias due to neglecting the rotational splitting is negligible if the rotational splitting remains lower than the resolution. When it is not the case, the bias is more important for narrow modes, leading to an overestimation of the width and an underestimation of the height. However, amplitudes are not biased, even when the fit is poor because of the anti-correlation between height and width.}
			
			 From our simulation, it is clear that our approach is not suited for stars whose rotational splitting is much greater than the resolution and which have high inclination angle ($\gtrsim 30^\circ$).

\section{Application to four \emph{Kepler} stars} \label{sec:5:0}

	In this study we have used short-cadence time series obtained by the {\it Kepler} mission during quarters 5 to 7. The corrected flux of each star has been computed following the procedures described by \cite{Garcia2011} to remove instrumental effects. The power spectral density has been obtained using a Lomb-Scargle algorithm and normalized following Parseval's theorem (frequency resolution of $\simeq 0.042$ $\mu$Hz).

\begin{figure}
\includegraphics*[angle=90,height=6.5cm]{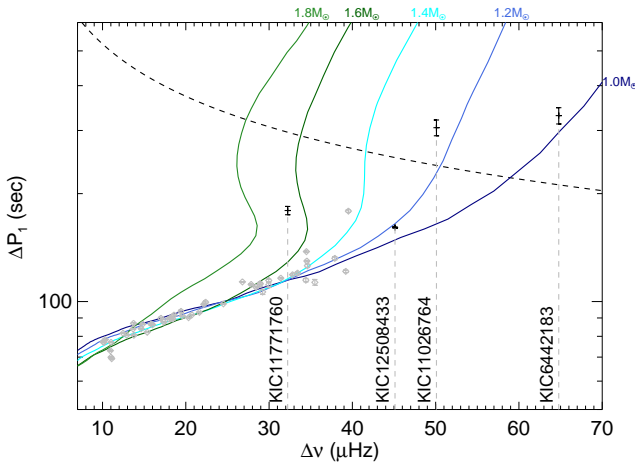}  
\caption{$\Delta{\Pi_1}-\Delta\nu$ diagram for the stars of the present study (black crosses) superimposed on the results from \cite{Mosser2012a} for red giant stars (gray diamonds). The dotted line indicates when $N_\pi$ = $N_\gamma$. Solid colored lines are models for masses between $1.0M_\odot$ and $1.8M_\odot$.}
\label{fig:Dp-Dnu}
\end{figure}

	We used three criteria in order to select stars to analyze. First, since the approach achieves its full potential when the splitting is not resolved or the inclination angle is small (cf. Section \ref{sec:4:1}), we kept only targets for which the splitting is not clearly visible in the power spectrum. Second, in order to have stars of various evolutionary stages and masses we selected targets of different large separation ($\Delta\nu$ from $32$ to $65$ $\mu$Hz) and mean period spacings (from $150$ to $300$ seconds). We kept the four stars with the highest signal-to-noise and carried out the analysis according to the method described Section \ref{sec:method}. This allowed us to measure all the individual properties of the modes (frequency, height, width, amplitude). Fig. \ref{fig:Dp-Dnu} shows the positions of these stars on a $\Delta{\Pi_1}-\Delta\nu$ diagram \citep{Mosser2012a} which gives us an indication on their evolutionary stage and their mass. 

\begin{figure*}
\includegraphics*[angle=90,height=6.5cm]{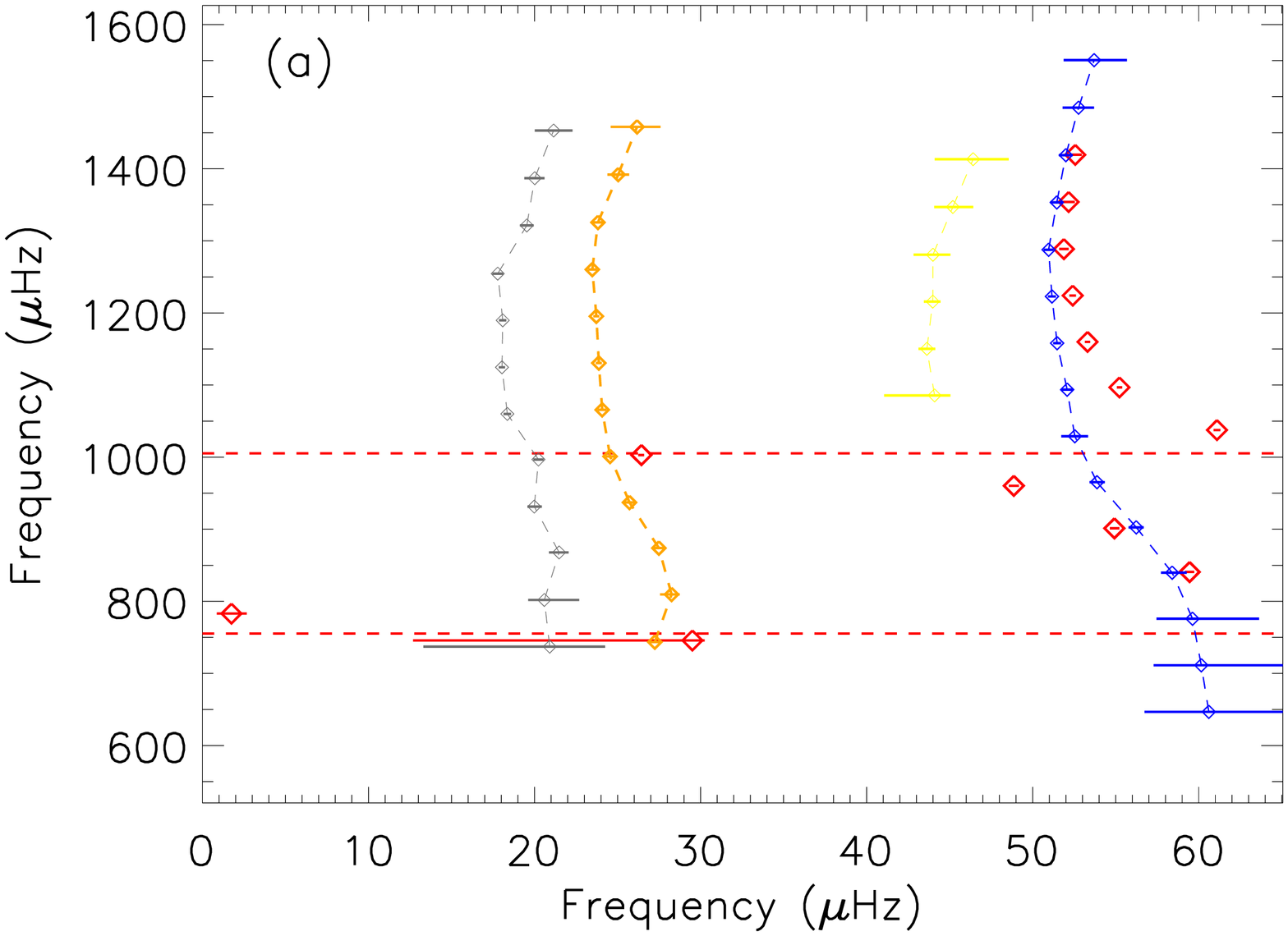}  
\includegraphics*[angle=90,totalheight=6.5cm]{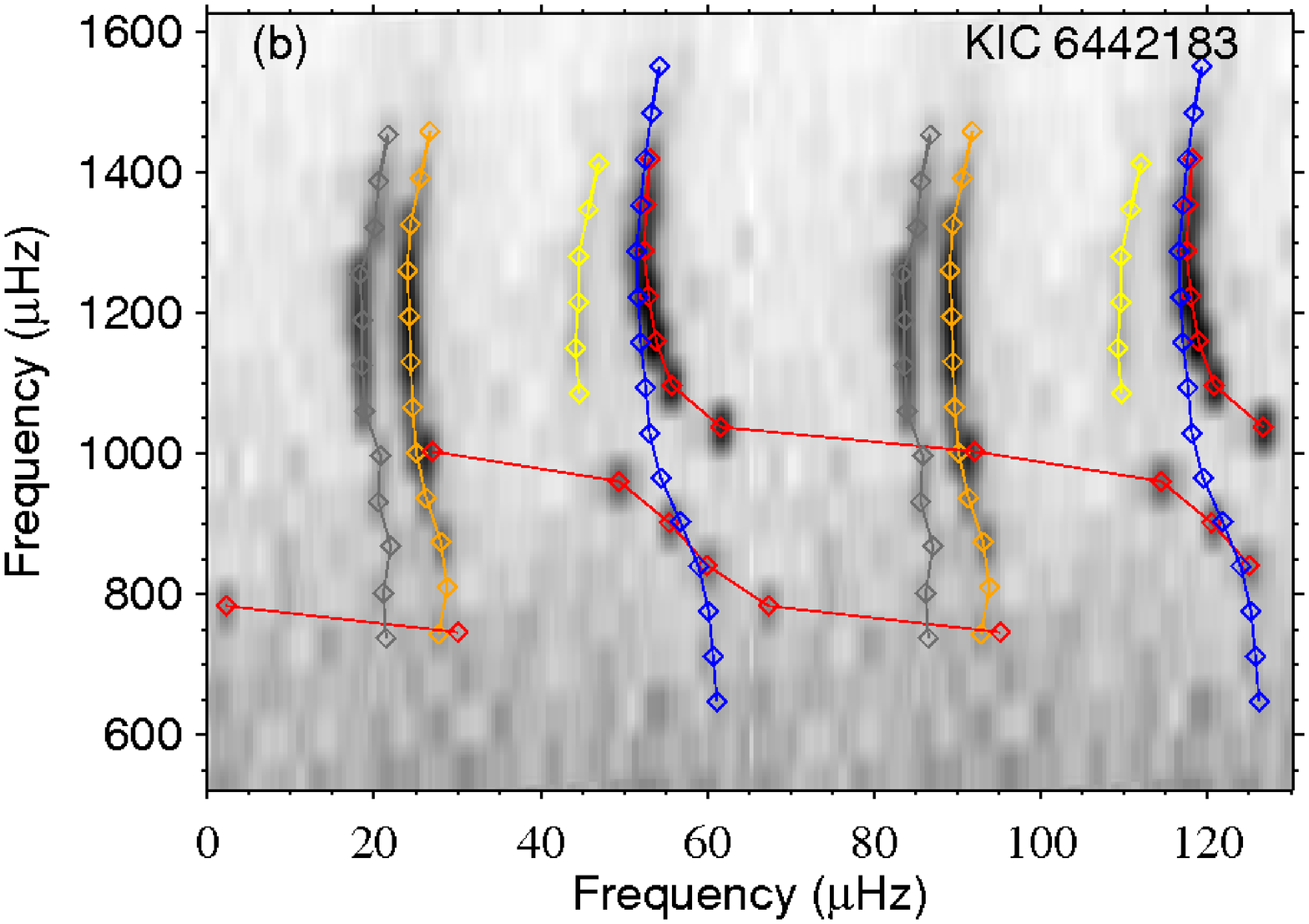} \\ 
\caption{\'Echelle diagrams of the fitted frequencies (a) and replicated \'echelle diagrams (b) for KIC 6442183.\textbf{Orange, gray, yellow colored dots represent $\ell=0,2,3$ pure p modes. $\ell=1$ fictitious $\pi$ modes are blue while $\ell=1$ observed mixed modes are red.} Horizontal dashed-red lines show where $\ell=1$ avoided crossing occur ($\gamma$-mode frequencies). All uncertainties are shown at $3 \sigma$.  }
\label{fig:Ech_diags1}
\end{figure*}

\begin{figure*}
\includegraphics*[angle=90,height=6.5cm]{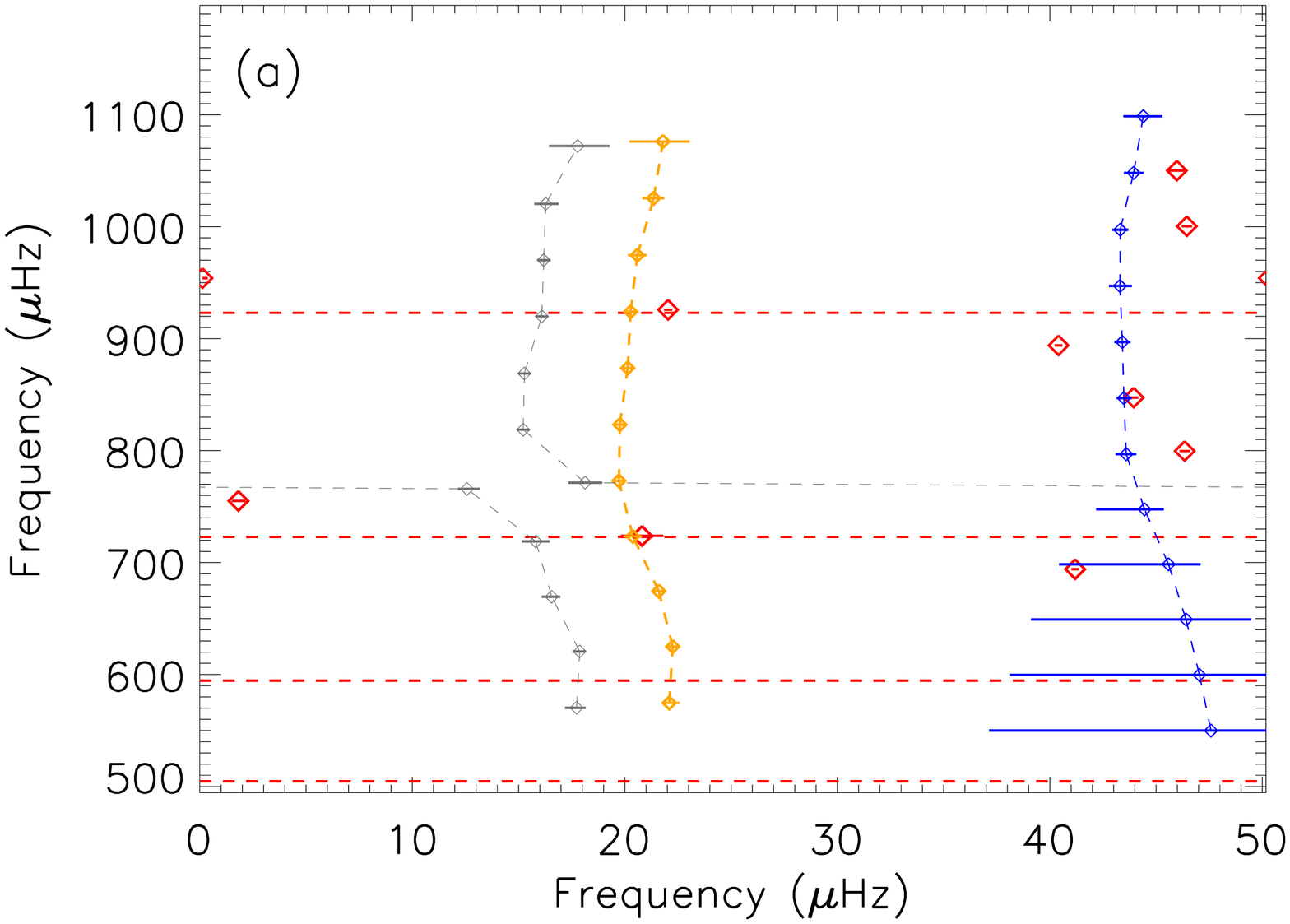} 
\includegraphics*[angle=90,totalheight=6.5cm]{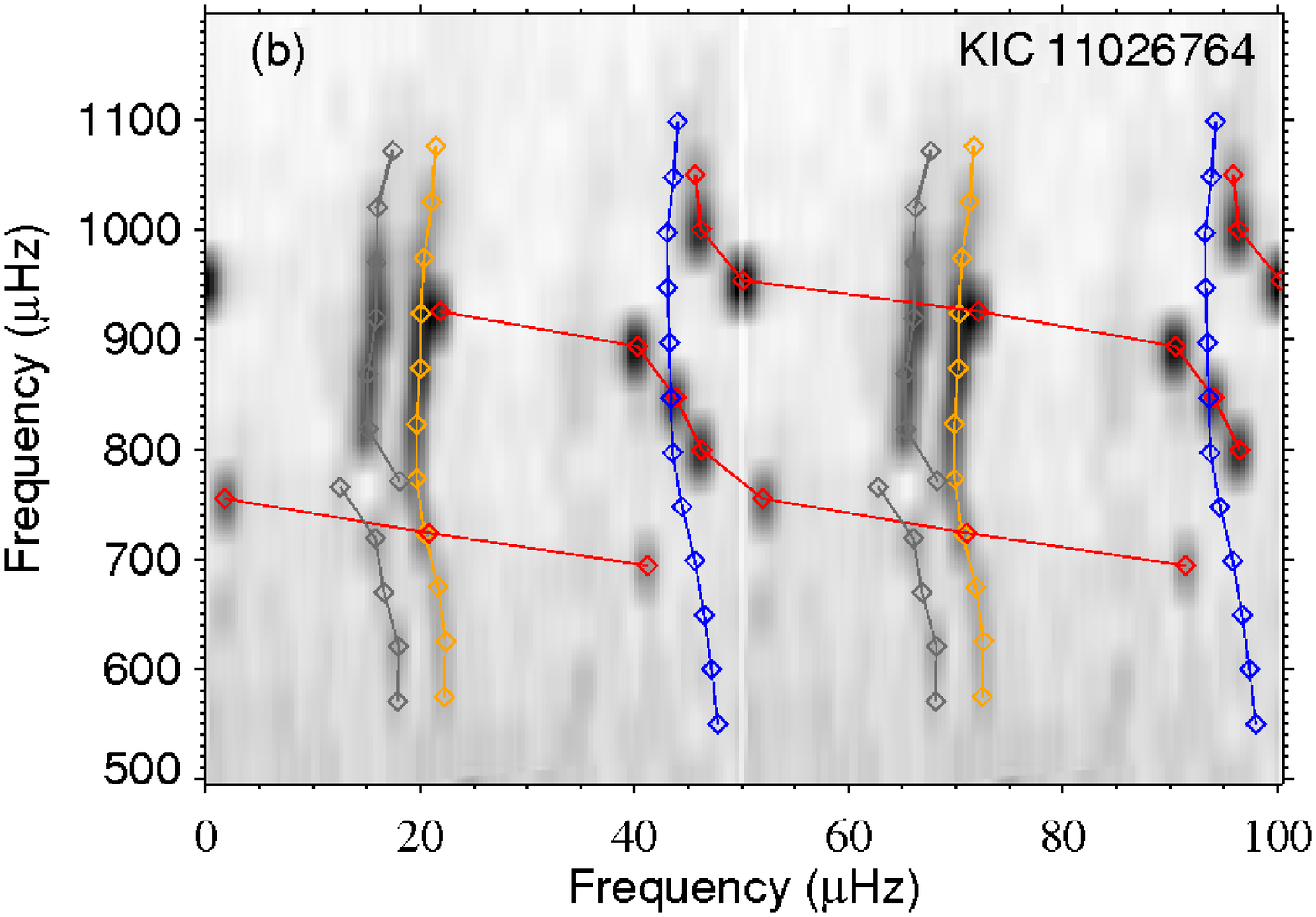} \\
\caption{\'Echelle diagrams of the fitted frequencies (a) and replicated \'echelle diagrams (b) for KIC 11026764. Same legend as Fig. \ref{fig:Ech_diags1}. }
\label{fig:Ech_diags2}
\end{figure*}

\begin{figure*}
\includegraphics*[angle=90,height=6.5cm]{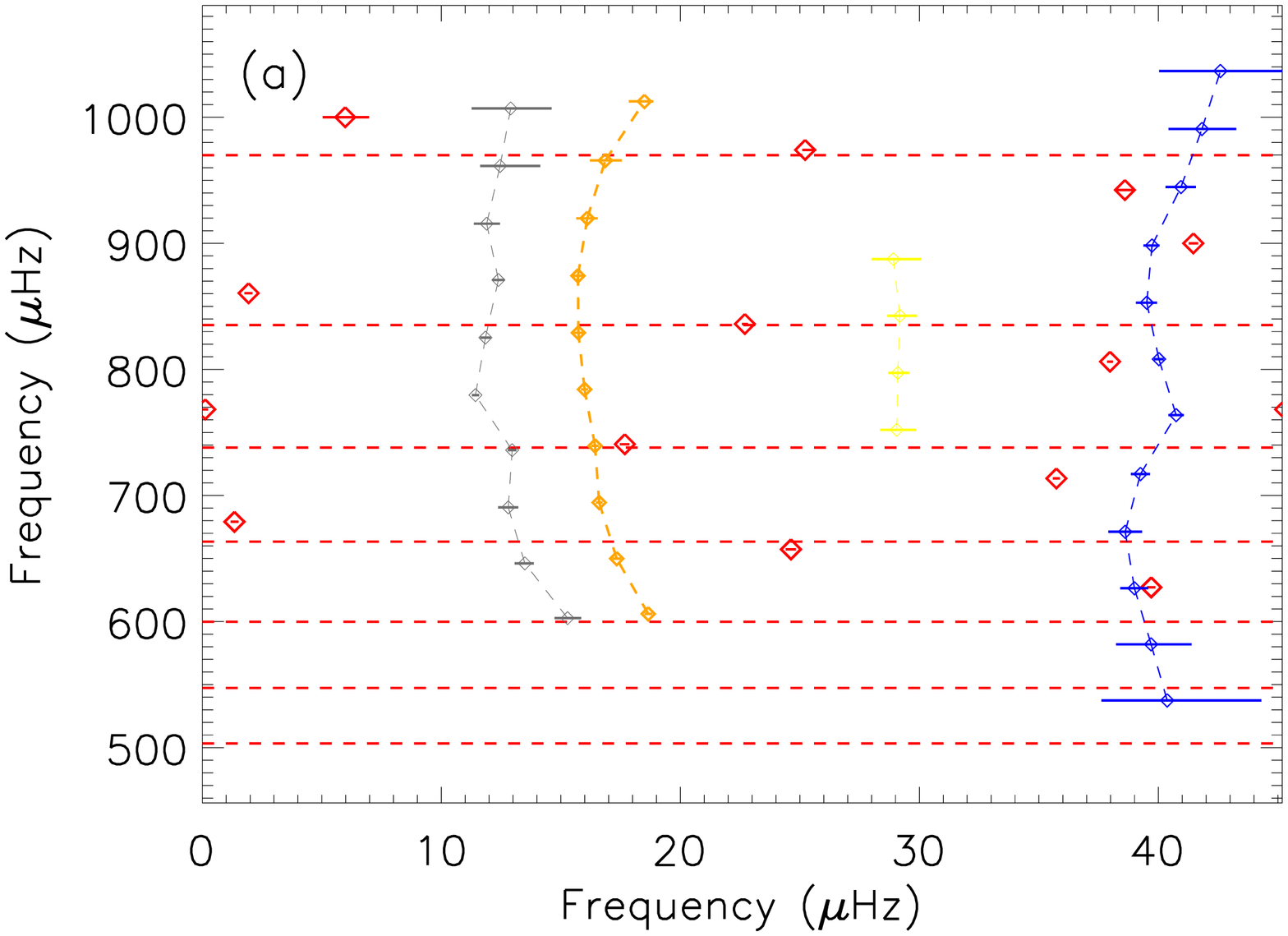} 
\includegraphics*[angle=90,totalheight=6.5cm]{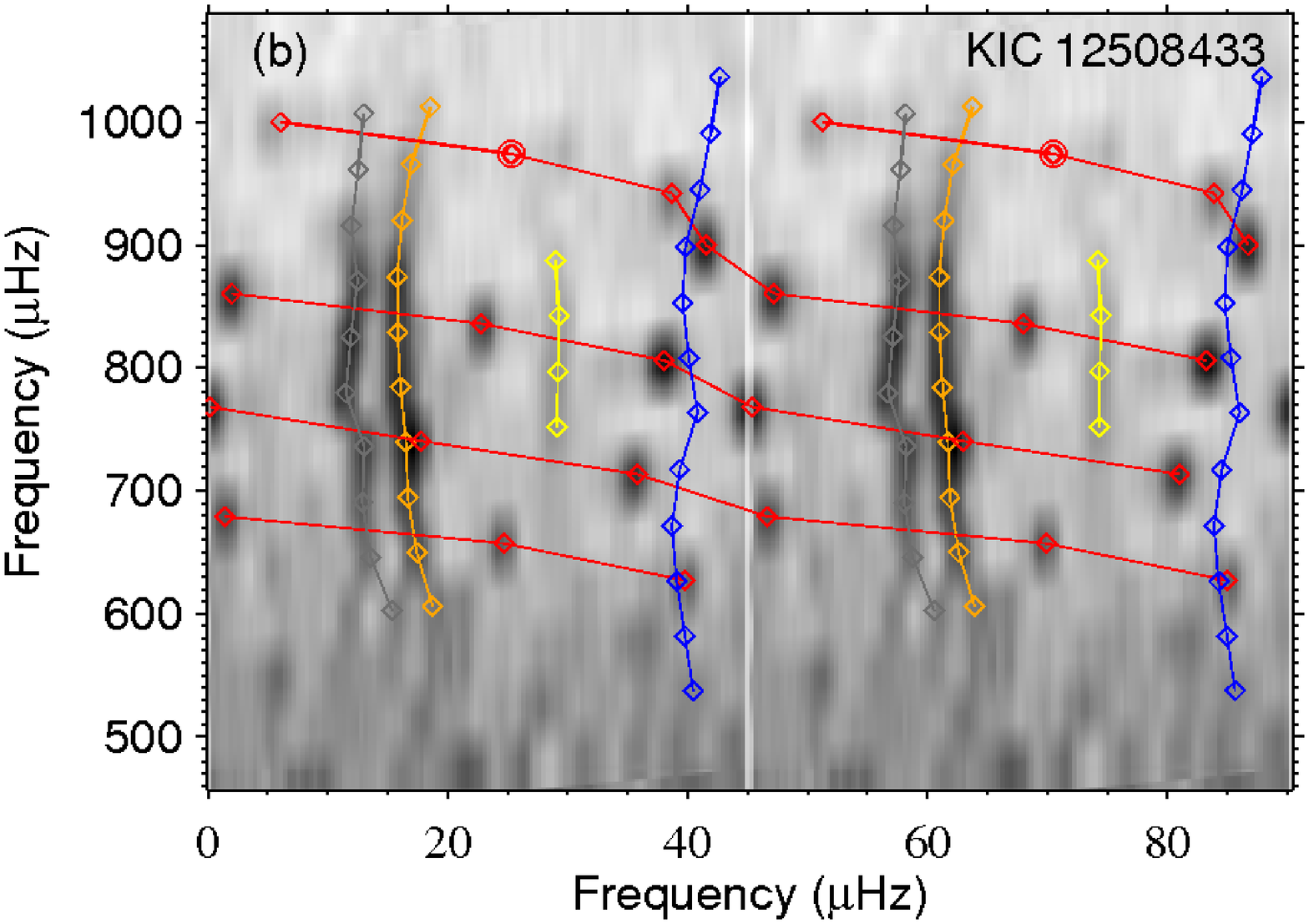}
\caption{\'Echelle diagrams of the fitted frequencies (a) and replicated \'echelle diagrams (b) for KIC 12508433. Same legend as Fig. \ref{fig:Ech_diags1}. The circle dot indicates the expected frequency of an $\ell=1$ mode but not statistically significant.} 
\label{fig:Ech_diags3}
\end{figure*}

\begin{figure*}
\includegraphics*[angle=90,height=6.5cm]{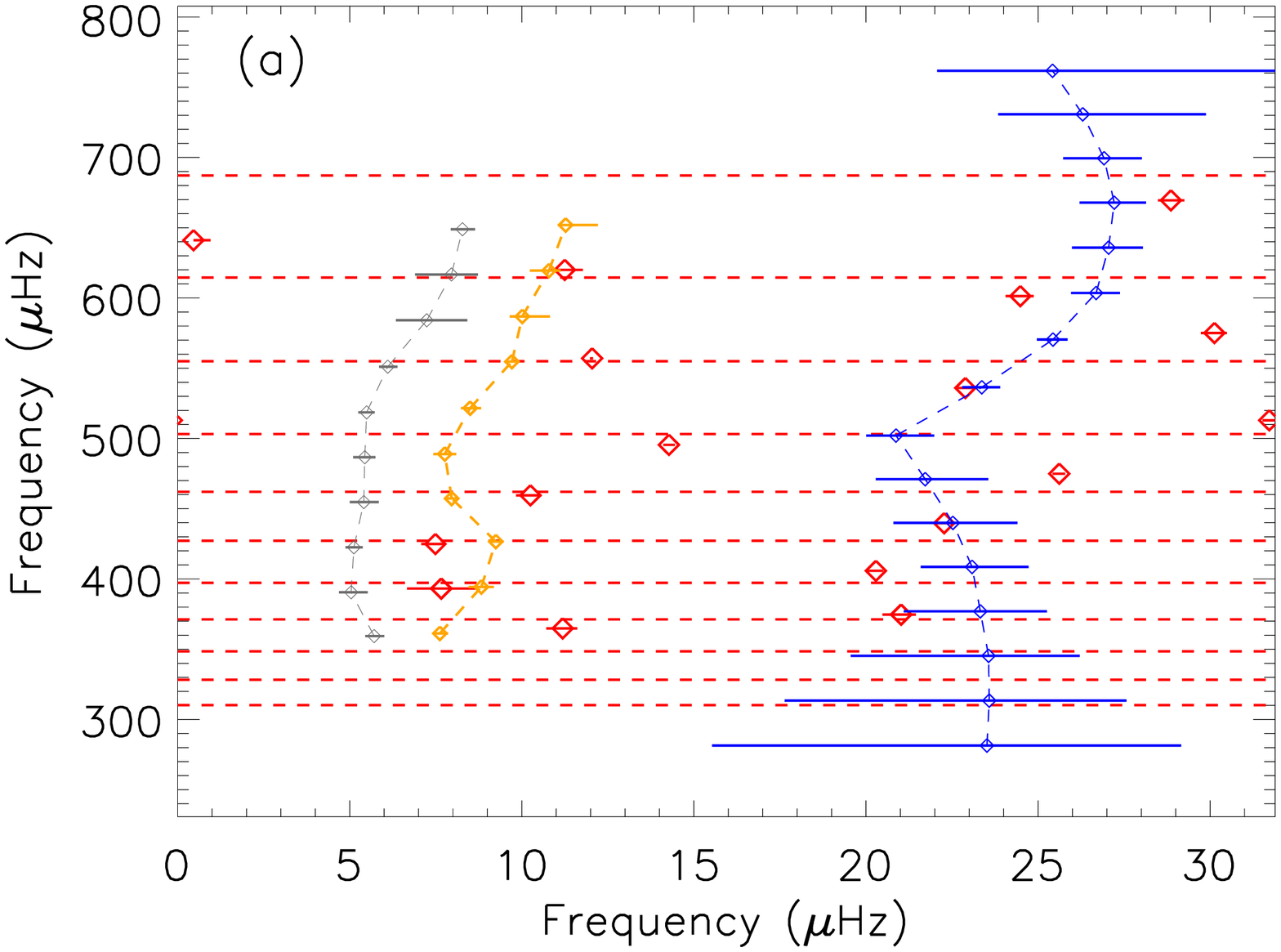} 
\includegraphics*[angle=90,totalheight=6.5cm]{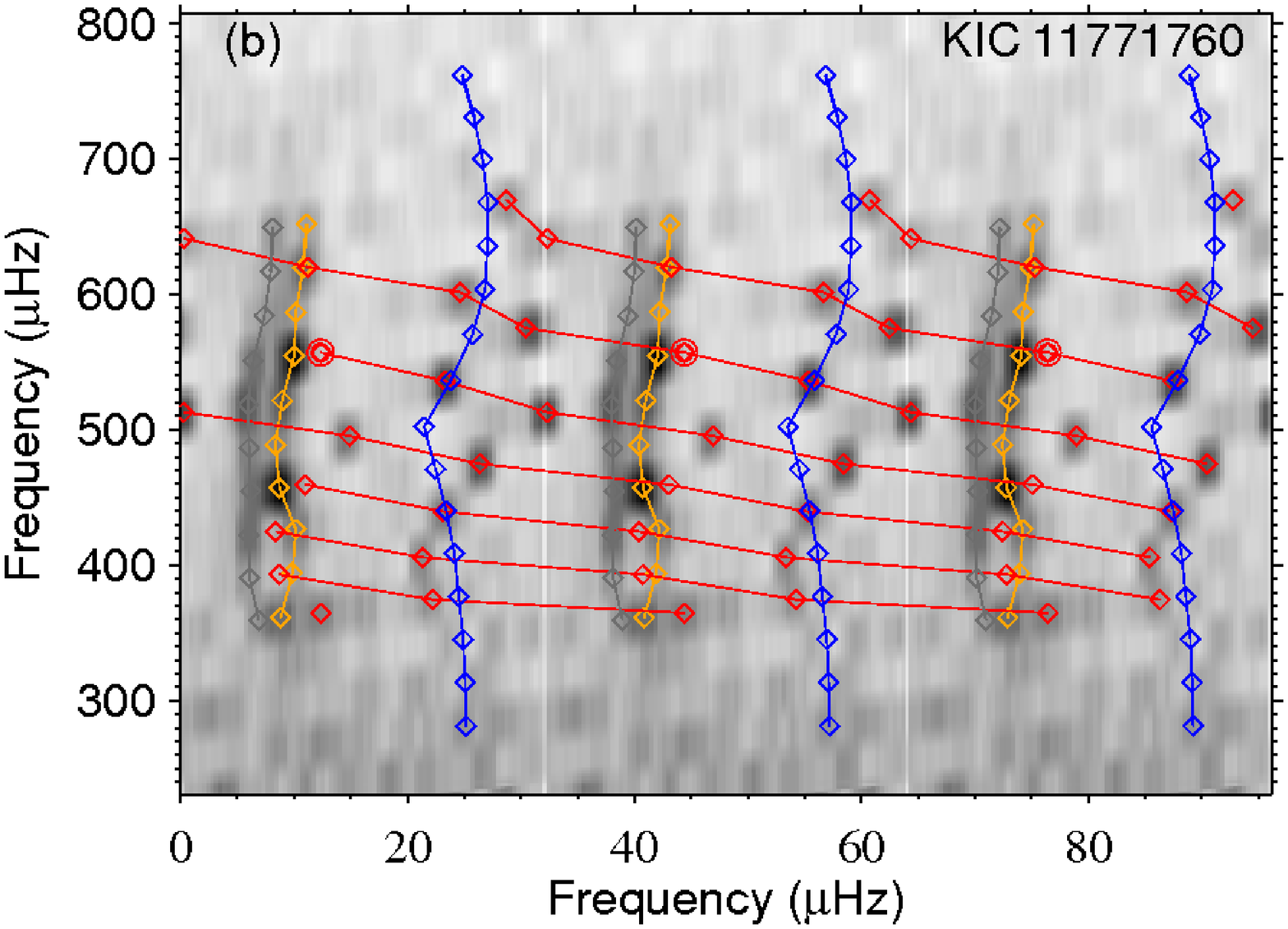}
\caption{\'Echelle diagrams of the fitted frequencies (a) and replicated \'echelle diagrams (b) for KIC 11771760. Same legend as Fig. \ref{fig:Ech_diags1}. The sharp feature on the $\ell=1$ $\pi$ modes is presumably an artifact. The circle dot indicates the expected frequency of an $\ell=1$ mode but not statistically significant.} 
\label{fig:Ech_diags4}
\end{figure*}

\subsection{Frequencies of the modes}

Table \ref{tab:parameters:mes} contains the main seismic frequency parameters, computed from the fit.  \textbf{Note that $\Delta\nu$, $\epsilon_p$, $\Delta\Pi_1$ and $\epsilon_{\gamma,1}$ are determined by fitting a first-order polynomial to the frequencies of $\ell=0$ p modes and $\ell=1$ $\gamma$ modes, at each iterative step of the MCMC process\footnote{\textbf{We therefore obtain the PDF for these asymptotic parameters from which we derive the median and the uncertainty.}}. Glitches inside stars introduce oscillations on $\Delta\nu$ and $\Delta\Pi_1$ that can slightly distort their determination, depending on the observed mode range. Therefore, to compare our values with theoretical $\Delta\nu$ and $\Delta\Pi_1$ one needs to use the same range of frequencies, as specified in the Tables \ref{tab:parameters:mes:all:KIC6442183}-\ref{tab:parameters:mes:all:KIC11771760}. These tables also contain the individual mode parameters of the analyzed stars.}

Scaling relations, were already discussed by \cite{Eddington1917} for Cepheids \citep{Belkacem2012}. However scaling relations have been introduced in the 80's by \cite{Ulrich1986} and \cite{Brown1991} before being used for solar-like oscillators \citep{Kjeldsen1995}. We used them to determine the masses and radii in Table \ref{tab:parameters:Stellar}. Note that scaling relations may suffer from systematics in some part of the HR diagram (such as for red clump star) because they are calibrated using the Sun \citep{Miglio2012}. Effective temperatures $T_\mathrm{eff}$ in Table \ref{tab:parameters:Stellar} are from \cite{Bruntt2012} and \cite{Pinsonneault2012}.

\begin{figure}
\includegraphics*[angle=90,height=6.5cm]{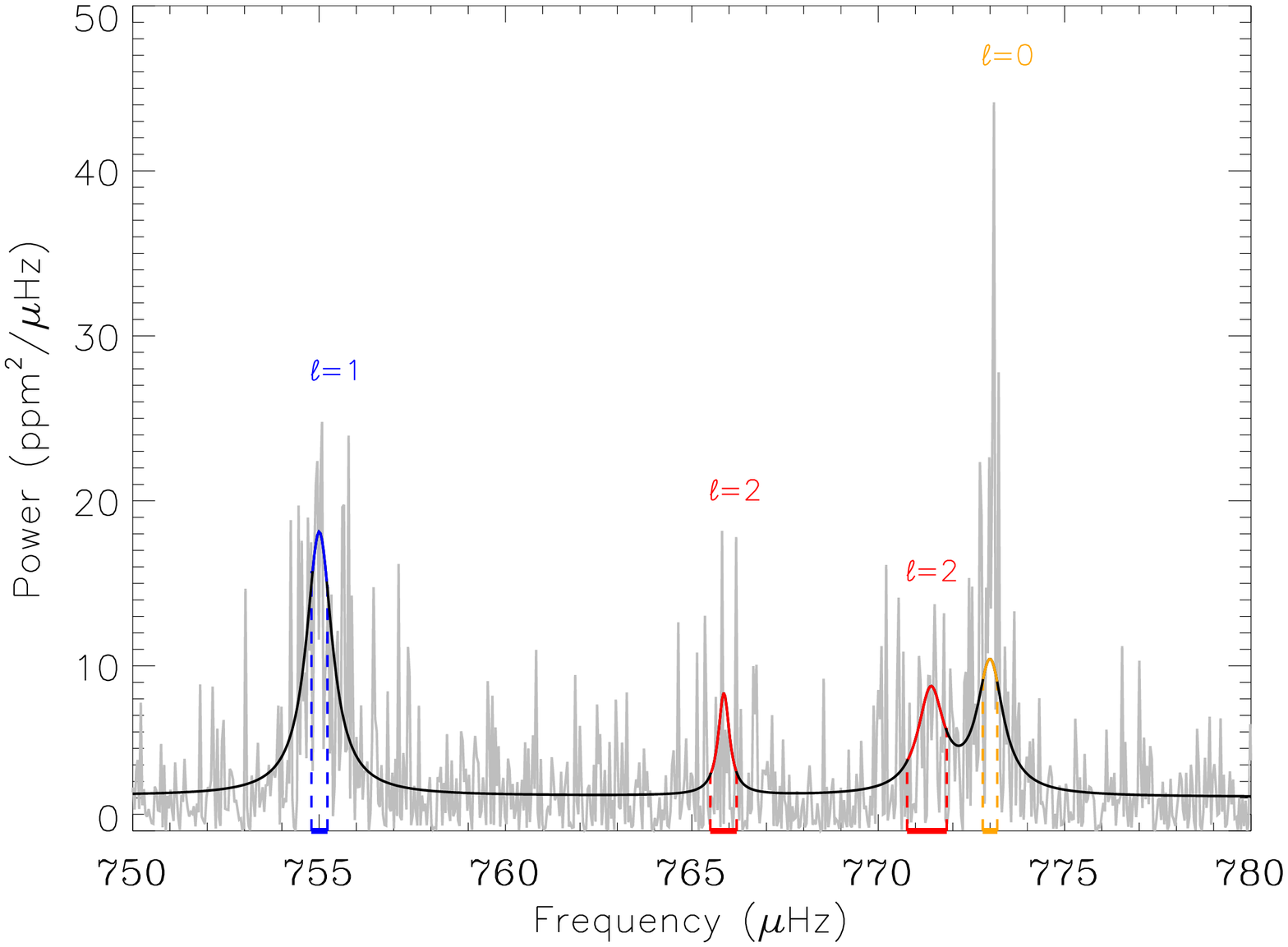}  
\caption{KIC 11026764 spectrum showing the $\ell=2$ mixed modes and the $\ell=2$ $\gamma$-mode ($\nu_{\gamma,2}$). Closest $\ell=0$ and $\ell=1$ modes are also represented. Thick-colored horizontal lines represent confidence intervals at $2\sigma$.}
\label{fig:Gemma-spec}
\end{figure}

\textbf{Figs. \ref{fig:Ech_diags1}-\ref{fig:Ech_diags4} show \'echelle diagrams (a) and replicated \'echelle diagrams (b) for the four stars. These diagrams show all fitted frequencies (p, $\pi$, $\gamma$ modes, and mixed modes). KIC 6442183 (Fig. \ref{fig:Ech_diags1}) and KIC 11026764 (Fig. \ref{fig:Ech_diags2}) have only a few avoided crossings and are therefore early subgiants. These avoided crossings can be seen as departures from the asymptotic relation for the p modes in the \'echelle diagram.} 

Interestingly, in addition to two clear $\ell=1$ avoided crossings, KIC 11026764 (also known as Gemma \emph{e.g.} \citealt{Metcalfe2010}), shows one clear $\ell=2$ avoided crossing (Fig. \ref{fig:Gemma-spec}), which was expected from models of this star but had not been observed before. Their potential interest is discussed Section \ref{sec:core_nature}. The coupled oscillator model extracts very precisely the position of the $\pi$ modes allowing an accurate determination of the frequency separation of each mixed modes from the $\pi$ modes home ridge, denoted $d$ in the following. This permits us to quantify the effect of the degree of mixture, on the individual modes parameters, as discussed below in Section \ref{sec:5:3}.

KIC 12508433 (Fig. \ref{fig:Ech_diags3}) and KIC 11771760 (Fig. \ref{fig:Ech_diags4}) are very interesting because contrary to KIC 6442183 and KIC 11026764, the density of $\gamma$ modes exceeds the density of $\pi$ modes, showing that these stars are more evolved. Their positions in the $\Delta{\Pi_1}$-$\Delta\nu$ diagram indicate that they are at the transition between the subgiant and giant phases \cite[ in prep.]{Mosser2012b}. 

Several $\ell=3$ modes are detected in KIC 6442183 and KIC 12508433. \textbf{These modes behave as pure p modes because the evanescent zone is larger}. Examination of the \'echelle diagram of KIC 11026764 with a strong smoothing, shows a slight power close to the $\pi$-mode ridge (Fig. \ref{fig:Ech_diags2}). These possible $\ell=3$ modes were not fitted because they are below the detection threshold with which we defined the initial guesses (cf. Section \ref{sec:method}, step 2). It is likely that more observations will allow an accurate fit to these $\ell=3$ modes, providing additional constraints to models.

\subsection{Testing the asymptotic relation for gravity modes using the $\gamma$ modes} \label{sec:core_nature}

Measuring gravity-modes of different degrees can greatly improve the constraints on the properties of stellar cores. From Eq.\ref{eq:ass_law2:1}, we can easily show that $\Delta{\Pi_{2}} = \Delta{\Pi_{1}} / \sqrt{3}$, which means that the density of $\ell=2$ modes is expected to be greater than for $\ell=1$. 
More interestingly, we also obtain,
\begin{equation} \label{eq:ass_law2:2}
	\epsilon_{\gamma, 2} - \epsilon_{\gamma, 1}= \left\{ 
		\begin{array}{ll}
				1/2 &\mbox{ (radiative core)} \\
			  0 &\mbox{ (convective core).}
		\end{array}
		\right.
\end{equation} %
Therefore, if both phase offsets for $\ell=1$ and $\ell=2$ $\gamma$ modes could be measured, their difference can only take two values, independent from the actual phase offset $\alpha_\gamma$. This removes the degeneracy on $\alpha_\gamma$ (see Section \ref{sec:priors:freq}) and potentially provides a direct insight on the nature of the core, independent to the stellar models. 

Subgiant and RGB stars should not have a convective core as Hydrogen is depleted in the core. Therefore convection is not required to transport energy and we can reasonably assume that $\epsilon_{\gamma, 2} - \epsilon_{\gamma,1} \simeq 1/2$. 
However, Eq.\ref{eq:ass_law2:1} is a first-order approximation and assumes that the radial order of the gravity modes is high. In early subgiants, the observed $\ell=1$ avoided crossings correspond to low order gravity modes. This implies that $\Delta \Pi_2$ may not scale with $\Delta \Pi_1 / \sqrt{3}$ and that $\alpha_{\gamma, \ell} = \alpha_{\gamma} + \ell/2$  may no longer be valid. Therefore, measuring $\epsilon_{\gamma, 2} - \epsilon_{\gamma, 1}$ allows us to probe the accuracy of the asymptotic relation for the gravity modes.

As discussed in Section \ref{sec:priors:others}, the mode bumping of $\ell=2$ mixed modes can hardly be observed as the coupling between the pressure and gravity modes is weaker than for $\ell=1$.
Nonetheless, among the four analyzed stars, KIC 11026764 and KIC 12508433 show clear $\ell=2$ bumped modes. In order to quantify approximately $\epsilon_{\gamma, 2}$, we performed a fit of the measured $\ell=2$ frequencies (output from the MCMC fit) in a similar fashion to \cite{Benomar2012b} and derived the expected frequencies of $\ell=2$ $\gamma$-mode $\nu_{\gamma, 2}$, causing the most significant mode bumping. 

For KIC 11026764, we obtained $\nu_{\gamma, 2} = 768.70 \pm 0.80$ $\mu$Hz. Using Eq.\ref{eq:ass_law2:1} and the measured $\Delta{\Pi_{1}}$ this leads to $\epsilon_{\gamma,2}=0.49 \pm 0.20$. We have $\epsilon_{\gamma,1}=0.59 \pm 0.20$ (see Table \ref{tab:parameters:mes}) which gives a difference $\epsilon_{\gamma, \ell=2} - \epsilon_{\gamma, \ell=1} = 0.10 \pm 0.20$. \textbf{This value is rather low and suggest that we are not in the asymptotic regime}\footnote{\textbf{Note that $n_{\gamma,1} =4$, which also suggests that we are not in the asymptotic regime.}}.

For KIC 12508433, which is more evolved, the main mode bumping is at $\nu_{\gamma, 2} = 737.74 \pm 4.50$ $\mu$Hz and we obtain $\epsilon_{\gamma,2}=0.91 \pm 0.02$ which leads to $\epsilon_{\gamma, \ell=2} - \epsilon_{\gamma, \ell=1} = 0.40 \pm 0.06$. This is more consistent with the asymptotic relation for the gravity modes than for KIC 11026764. 

\subsection{Heights, widths and amplitudes of the modes} \label{sec:5:3}

While pulsation frequencies provide constraints on the mass, radius and evolutionary stage of stars, the widths, heights and amplitudes of the modes may carry information on \textbf{the mode inertia and on the non-adiabatic effects such as the radiative damping.} Measuring them is therefore essential to constrain stellar models and to better understand the role of convection in the excitation mechanism of the modes.

Due to the long observations achieved with \emph{Kepler}, many mixed modes are well resolved and an accurate fit is possible. Modes \emph{a posteriori} not statistically significant are discarded in the following analysis. 

\textbf{It is well known that width and height of modes are anti-correlated (e.g. \citealt{appourchaux98, benomar09}). This anti-correlation greatly reduces the precision of their measure, whereas their product, which defines the mode amplitude, is much better constrained. The relationship between height, width and amplitude is well defined, therefore in the following we will focus mainly on width and amplitude, without loss of information.}

\begin{figure*}
\begin{center}
\includegraphics[angle=90,totalheight=6.cm]{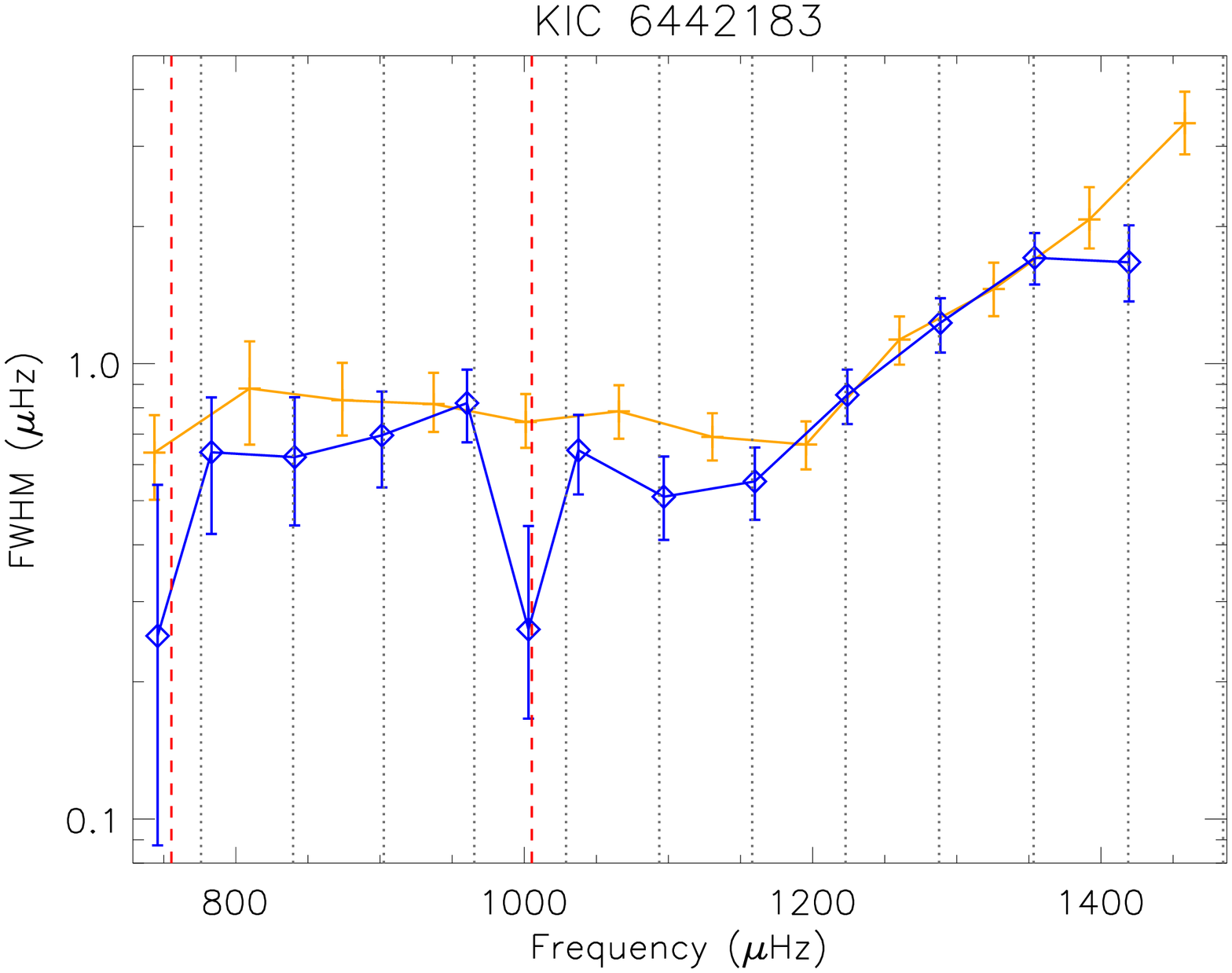} 
\includegraphics[angle=90,totalheight=6.cm]{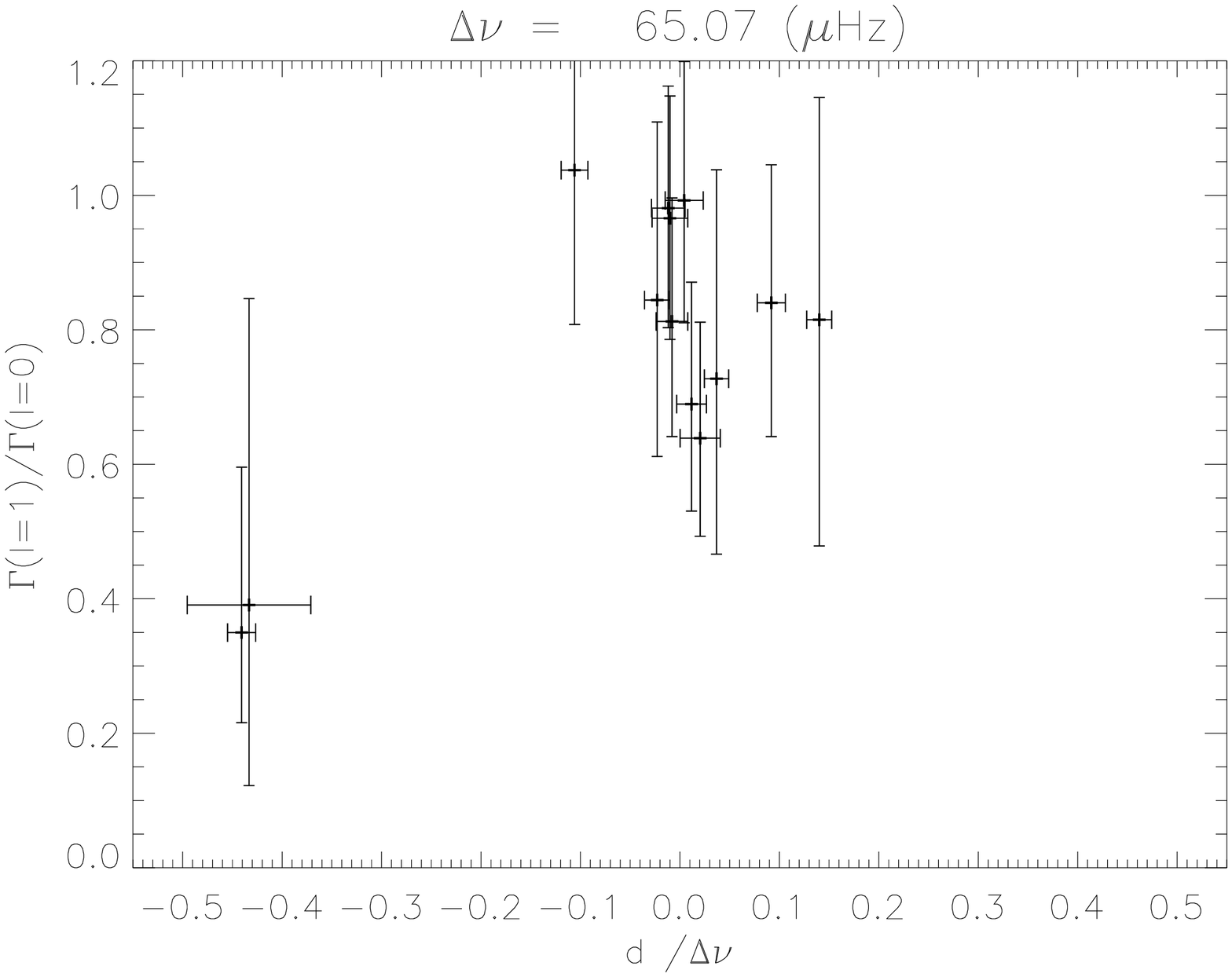}  \\ 
\includegraphics[angle=90,totalheight=6.cm]{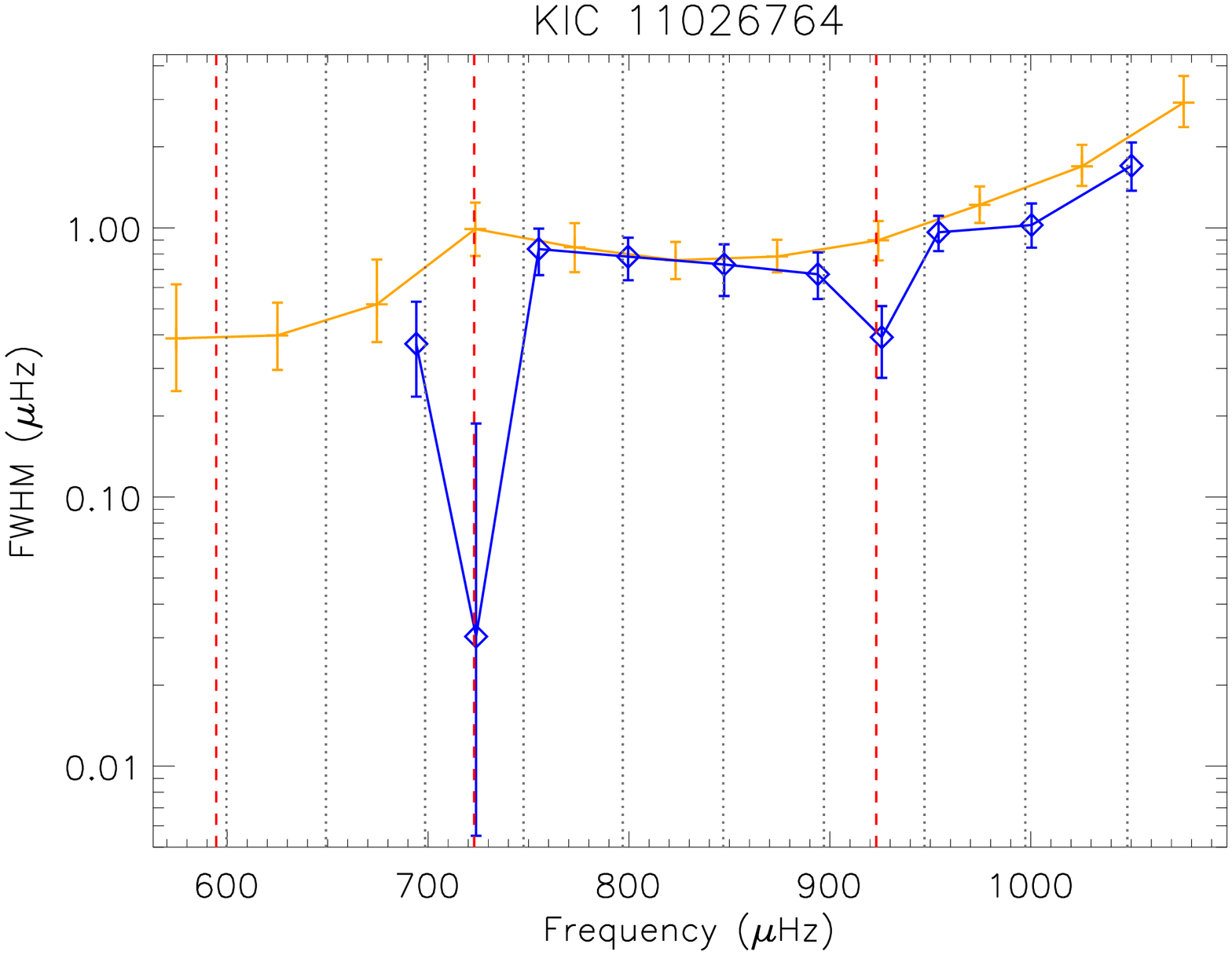}
\includegraphics[angle=90,totalheight=6.cm]{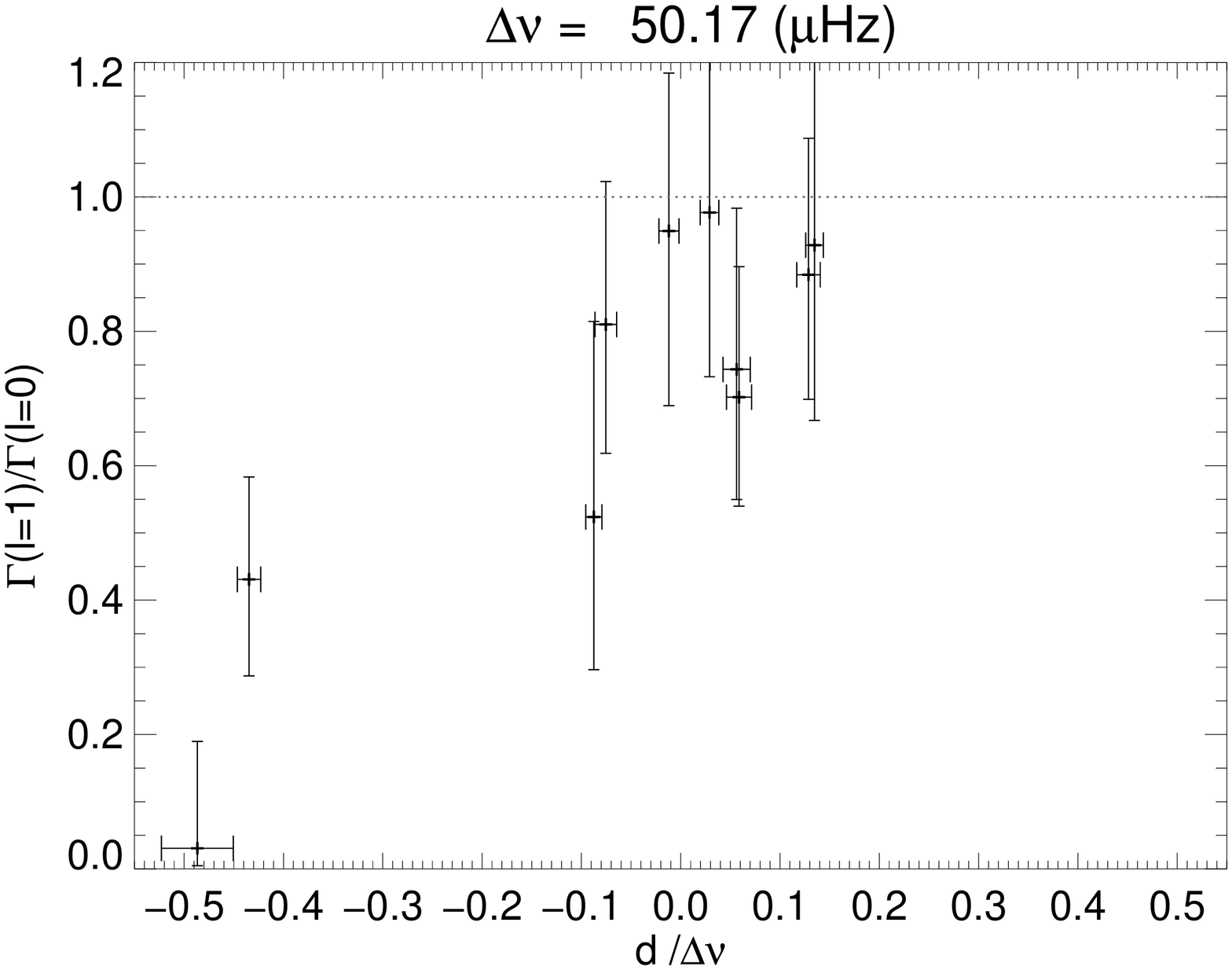} \\
\includegraphics[angle=90,totalheight=6.cm]{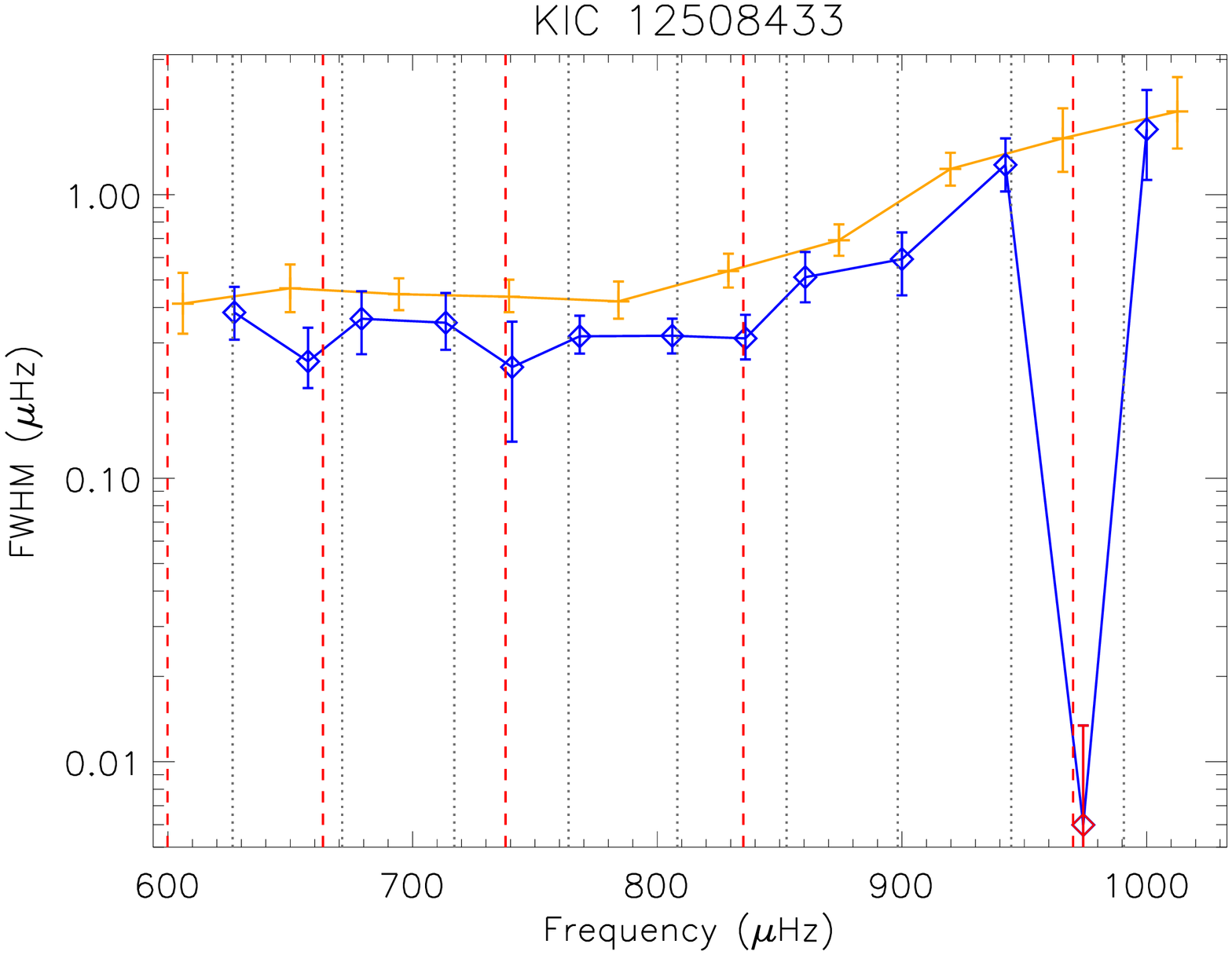} 
\includegraphics[angle=90,totalheight=6.cm]{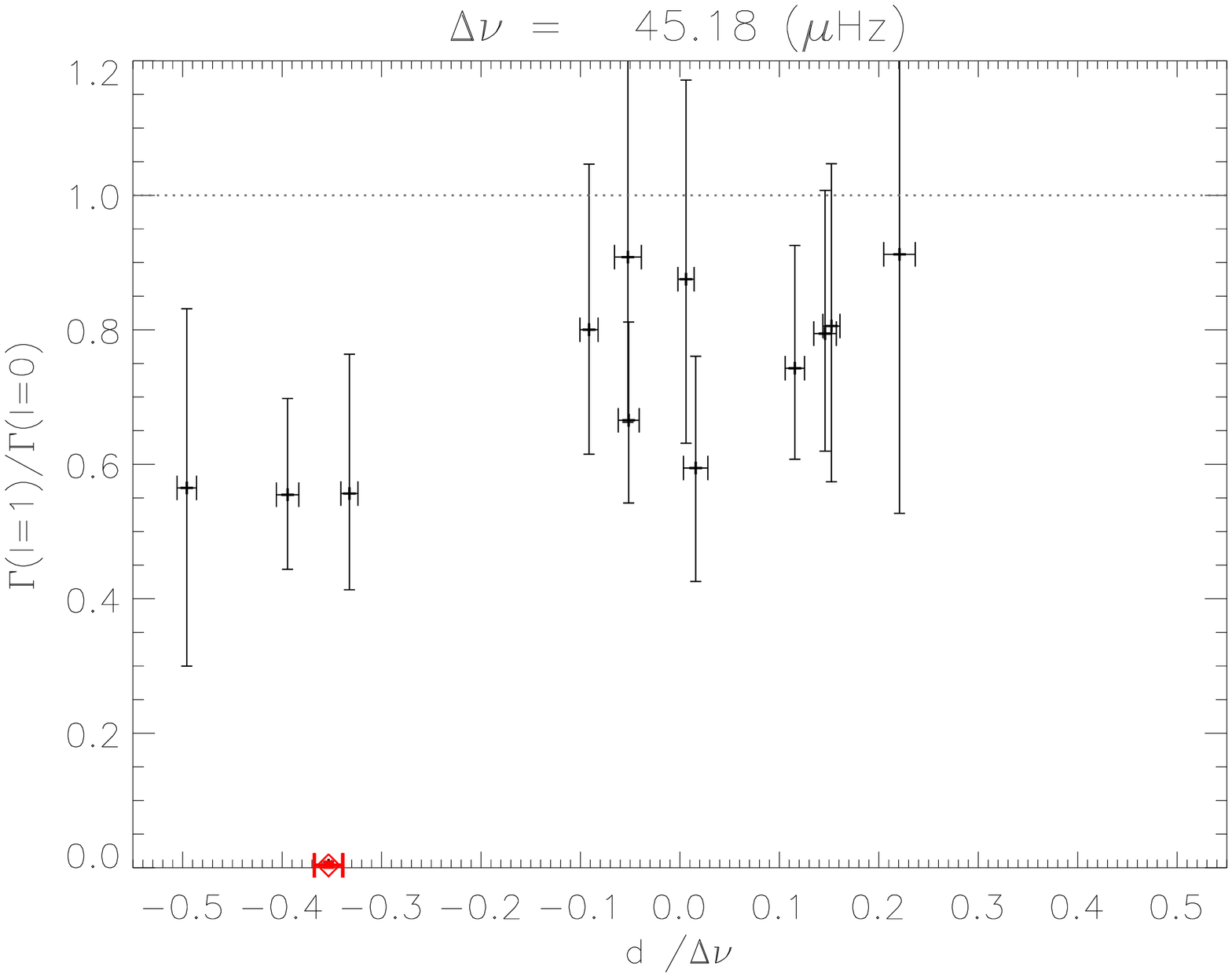} \\
\includegraphics[angle=90,totalheight=6.cm]{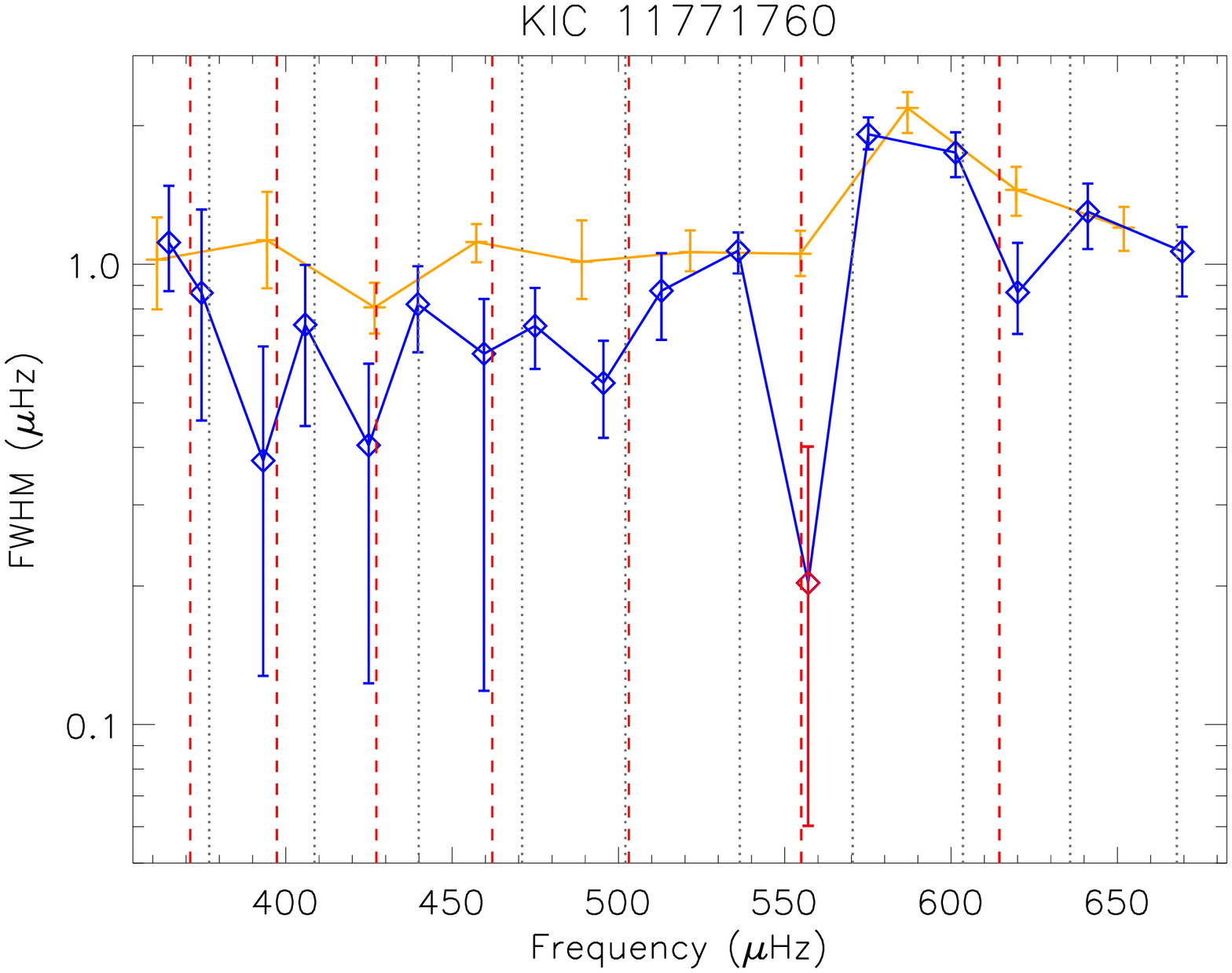}
\includegraphics[angle=90,totalheight=6.cm]{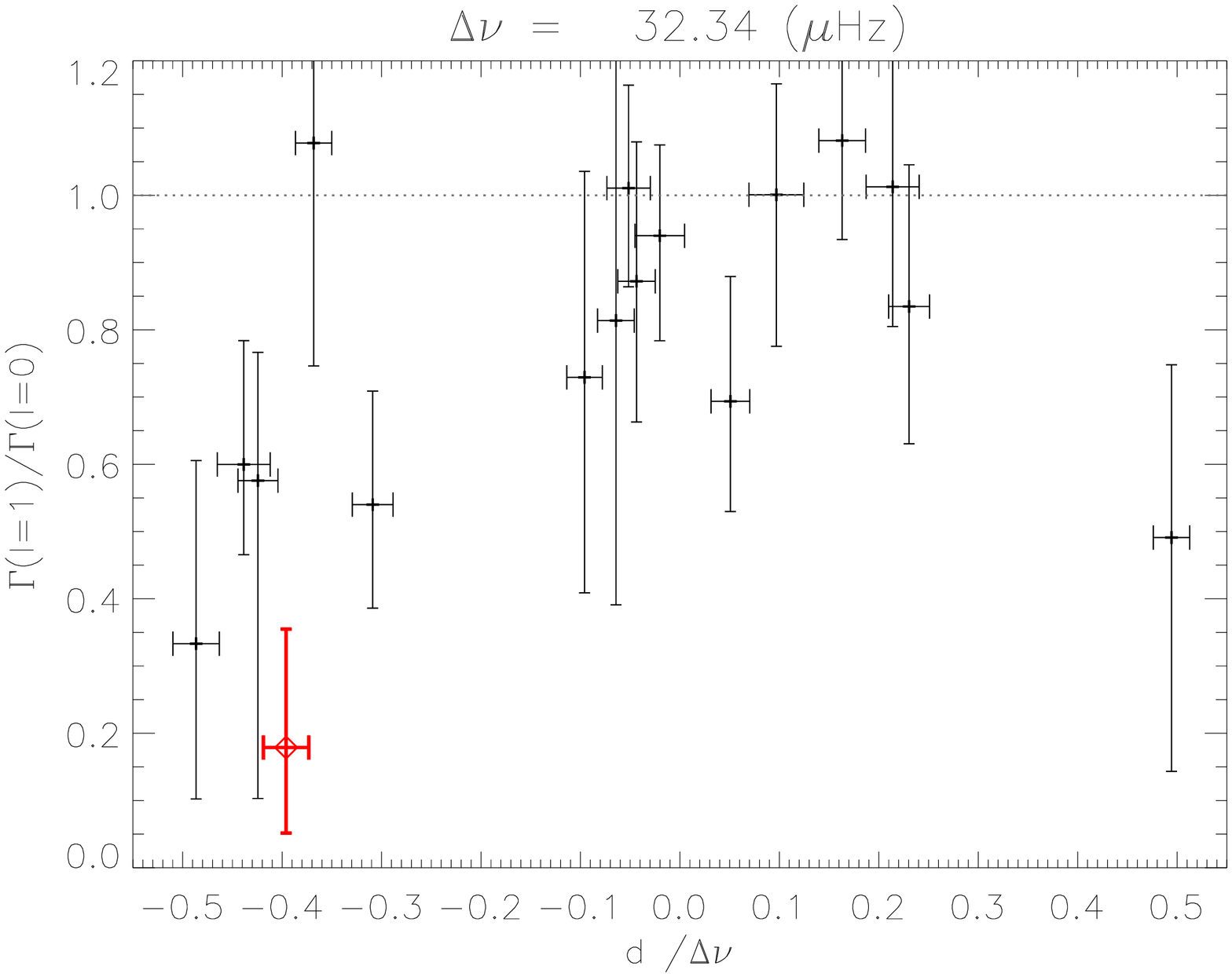} \\
\caption{$\ell=0$ (orange) and $\ell=1$ (blue) widths (FWHM, $\Gamma$). Vertical red dashed lines indicate $\gamma$ modes positions. Vertical black dotted lines indicate $\pi$ modes positions. Red data points are not statistically significant in the power spectrum. When the $\ell=1$ mode is far from the $\pi$ mode frequencies, width drops.}
\label{fig:Widths}
\end{center}
\end{figure*}

\begin{figure*}
\begin{center}
\includegraphics[angle=90,totalheight=6.cm]{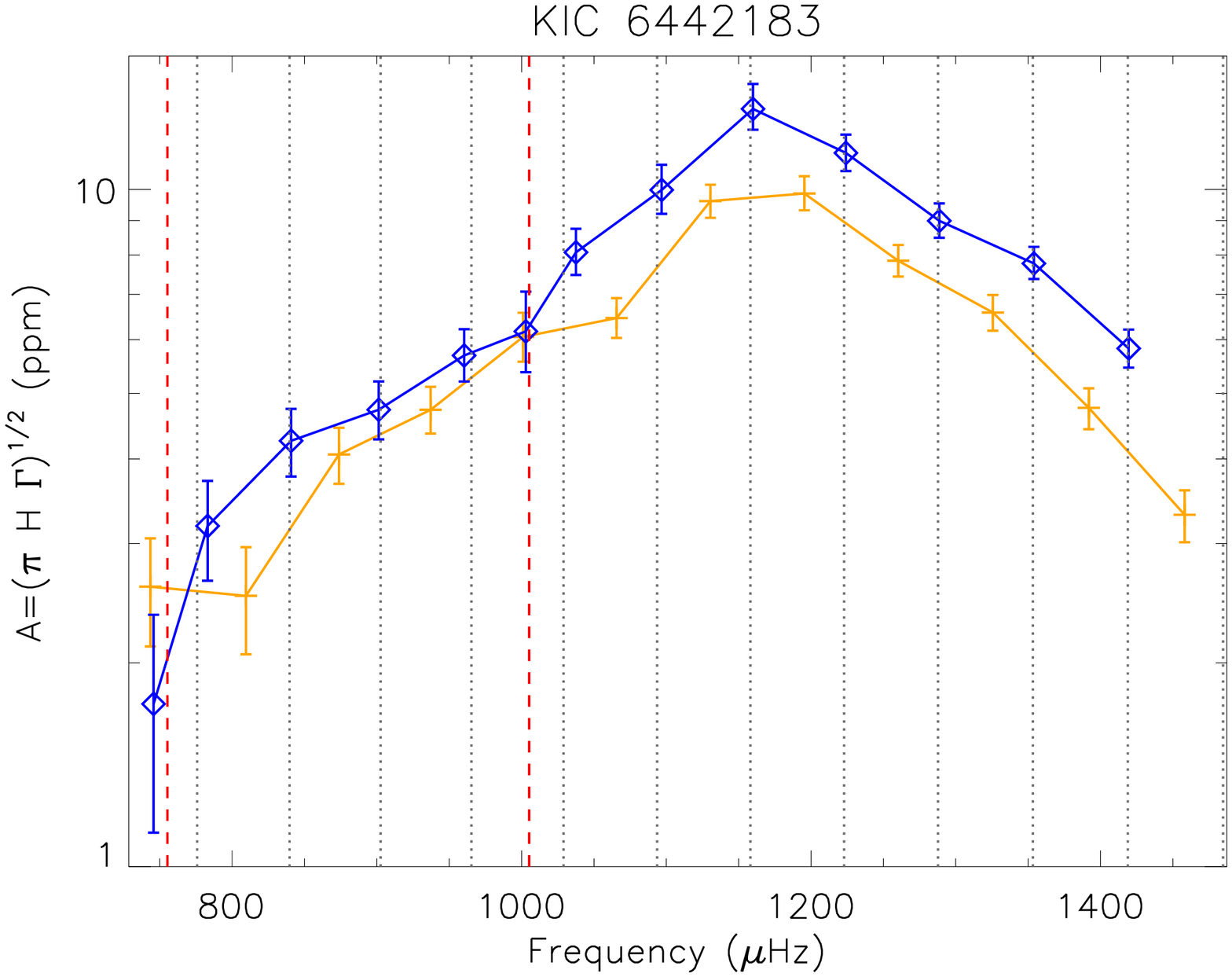} 
\includegraphics[angle=90,totalheight=6.cm]{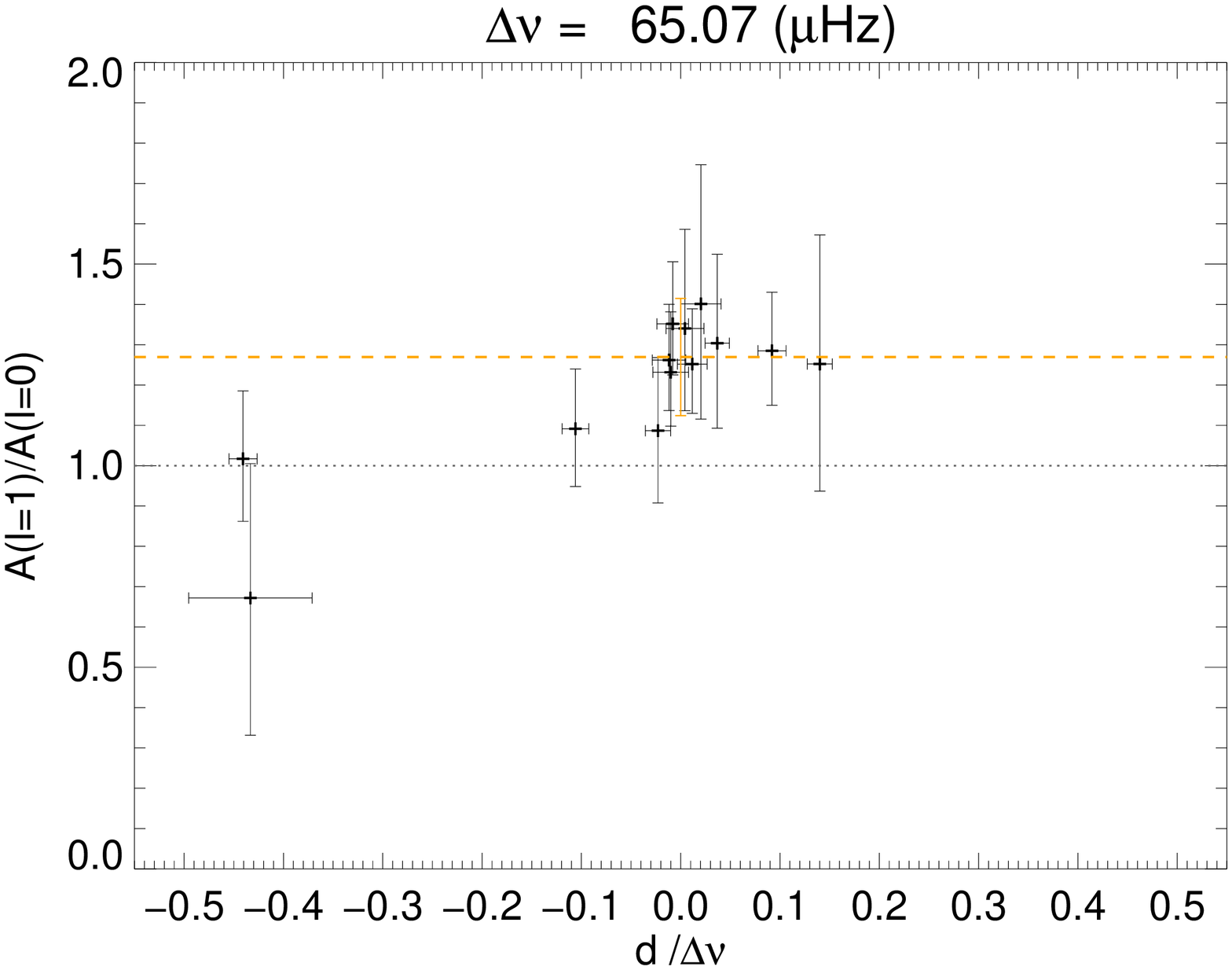}  \\ 
\includegraphics[angle=90,totalheight=6.cm]{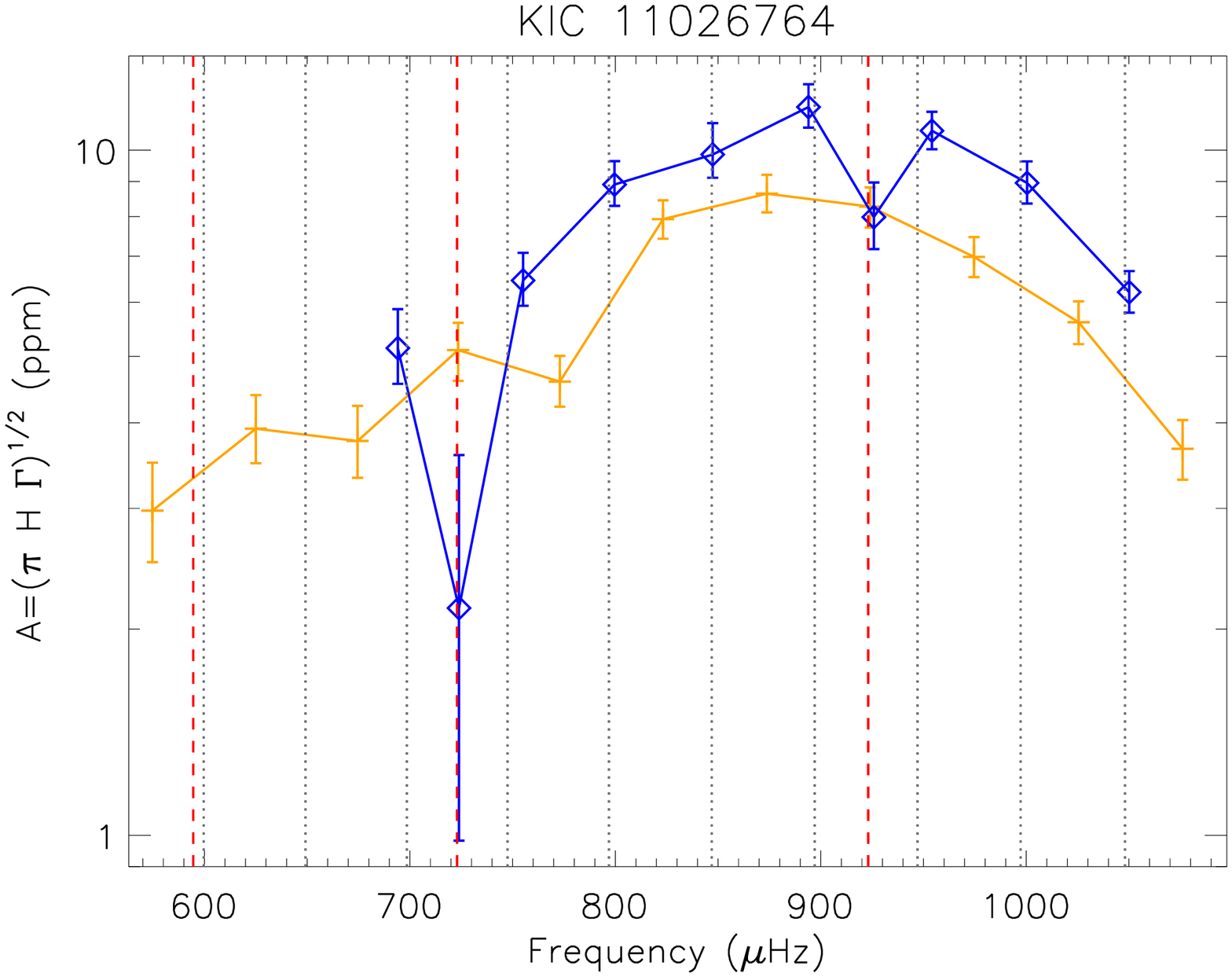}
\includegraphics[angle=90,totalheight=6.cm]{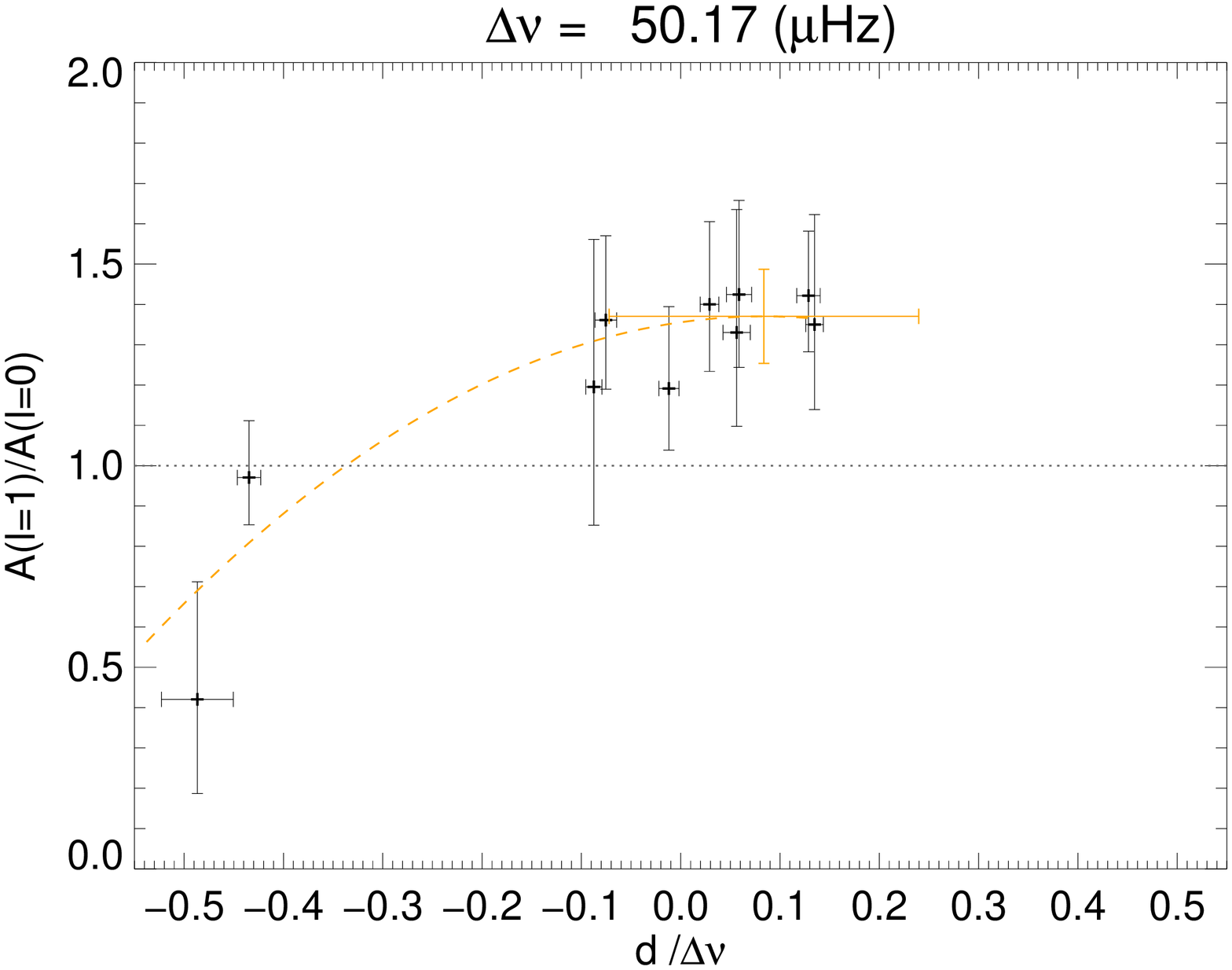} \\
\includegraphics[angle=90,totalheight=6.cm]{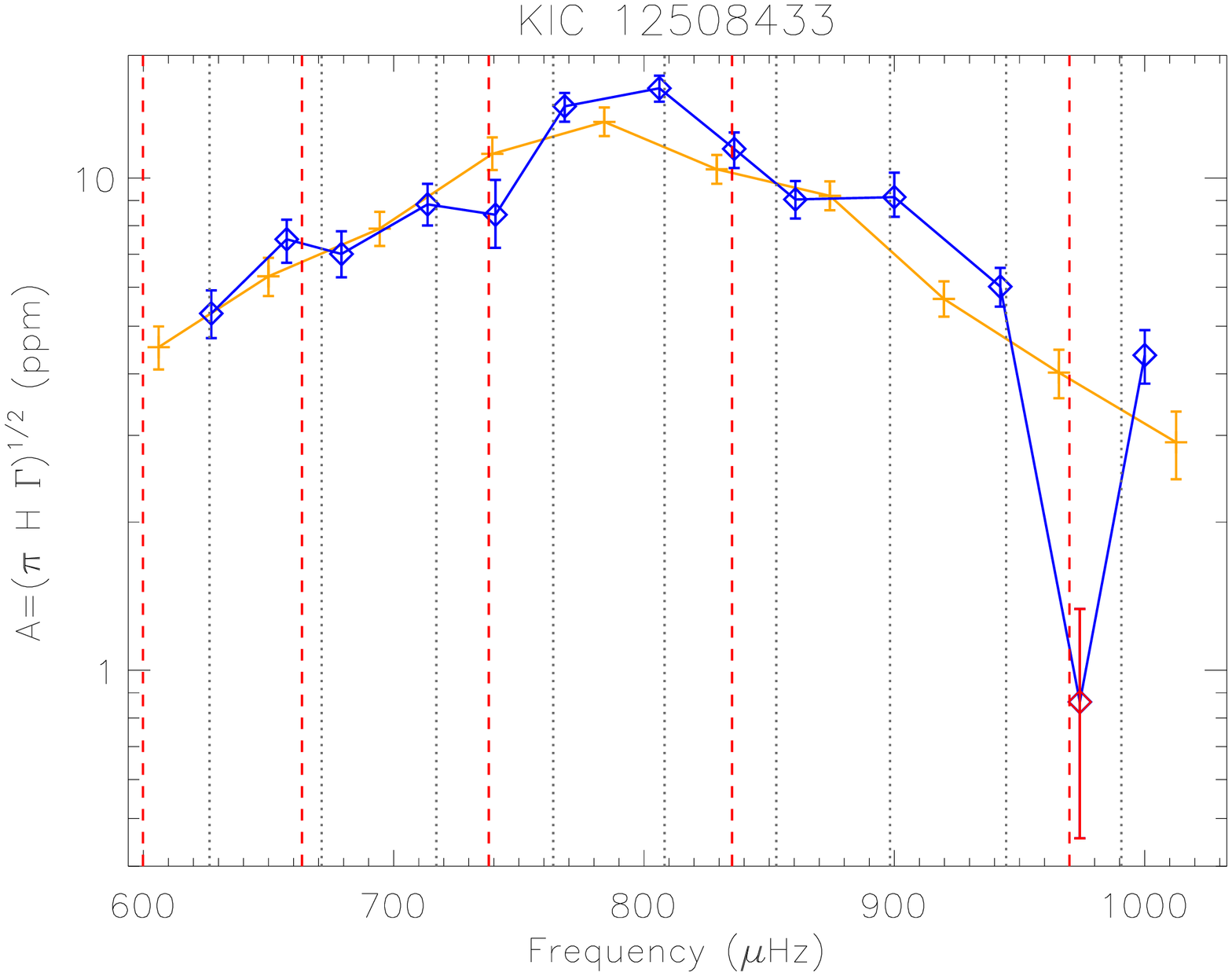} 
\includegraphics[angle=90,totalheight=6.cm]{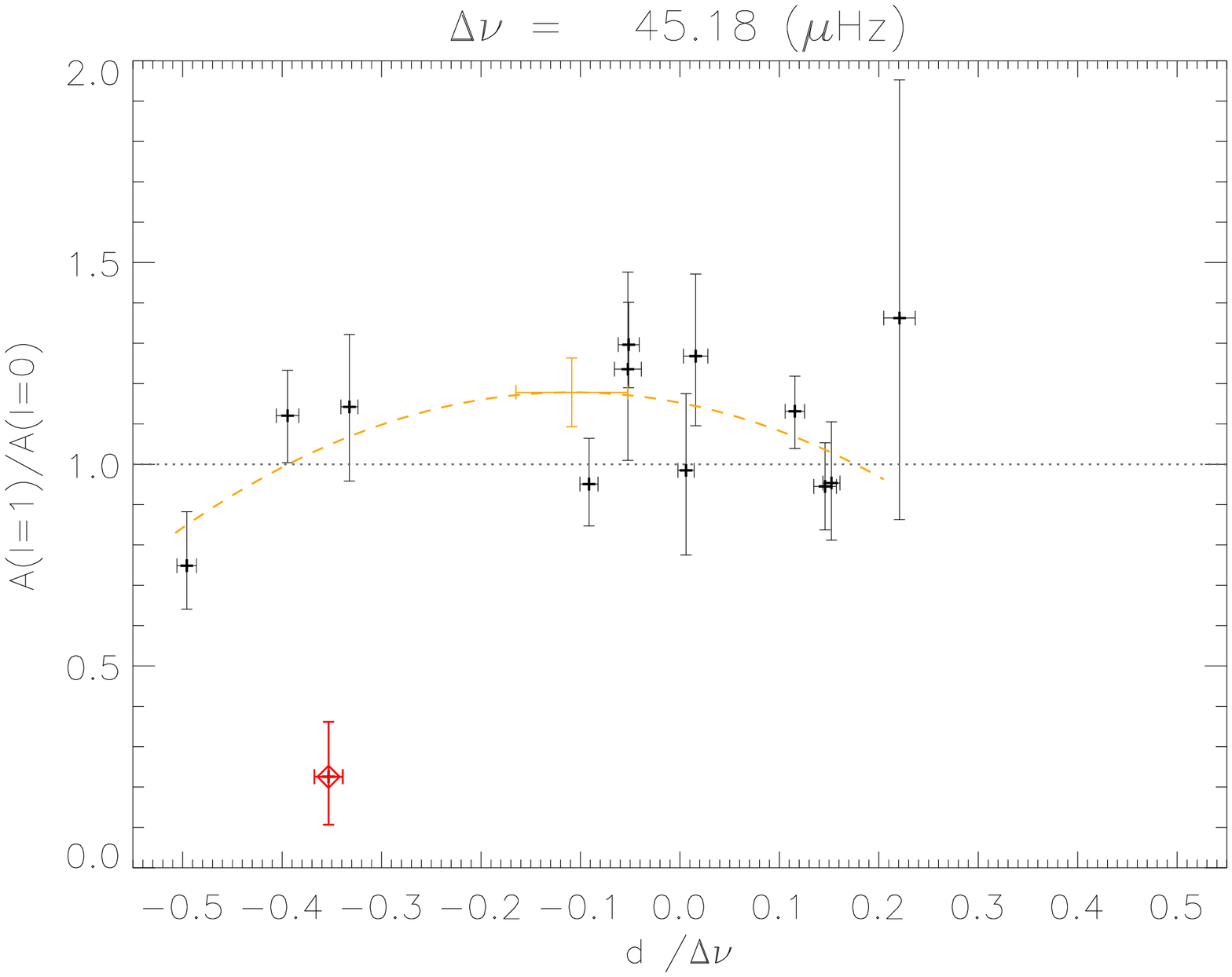} \\
\includegraphics[angle=90,totalheight=6.cm]{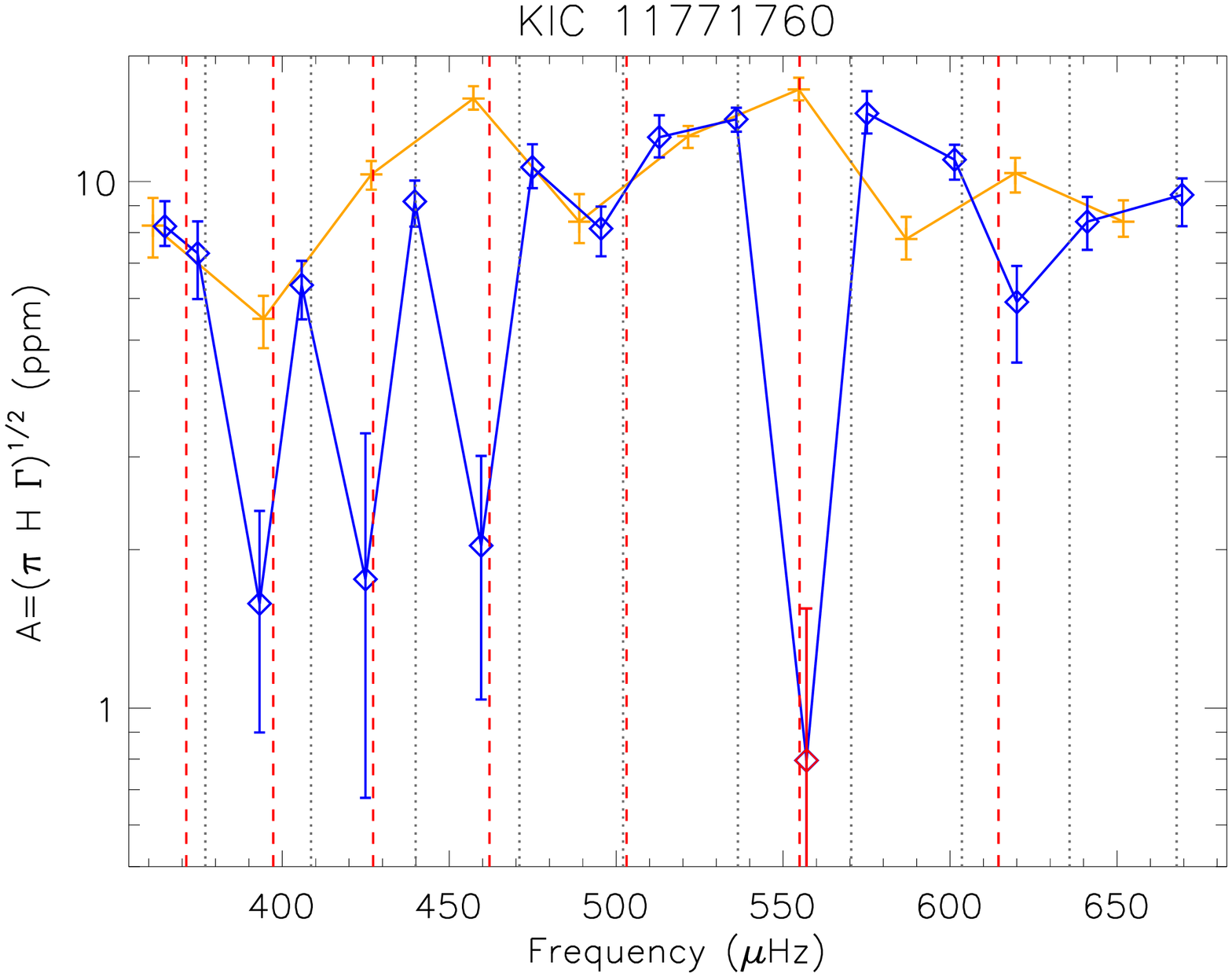}
\includegraphics[angle=90,totalheight=6.cm]{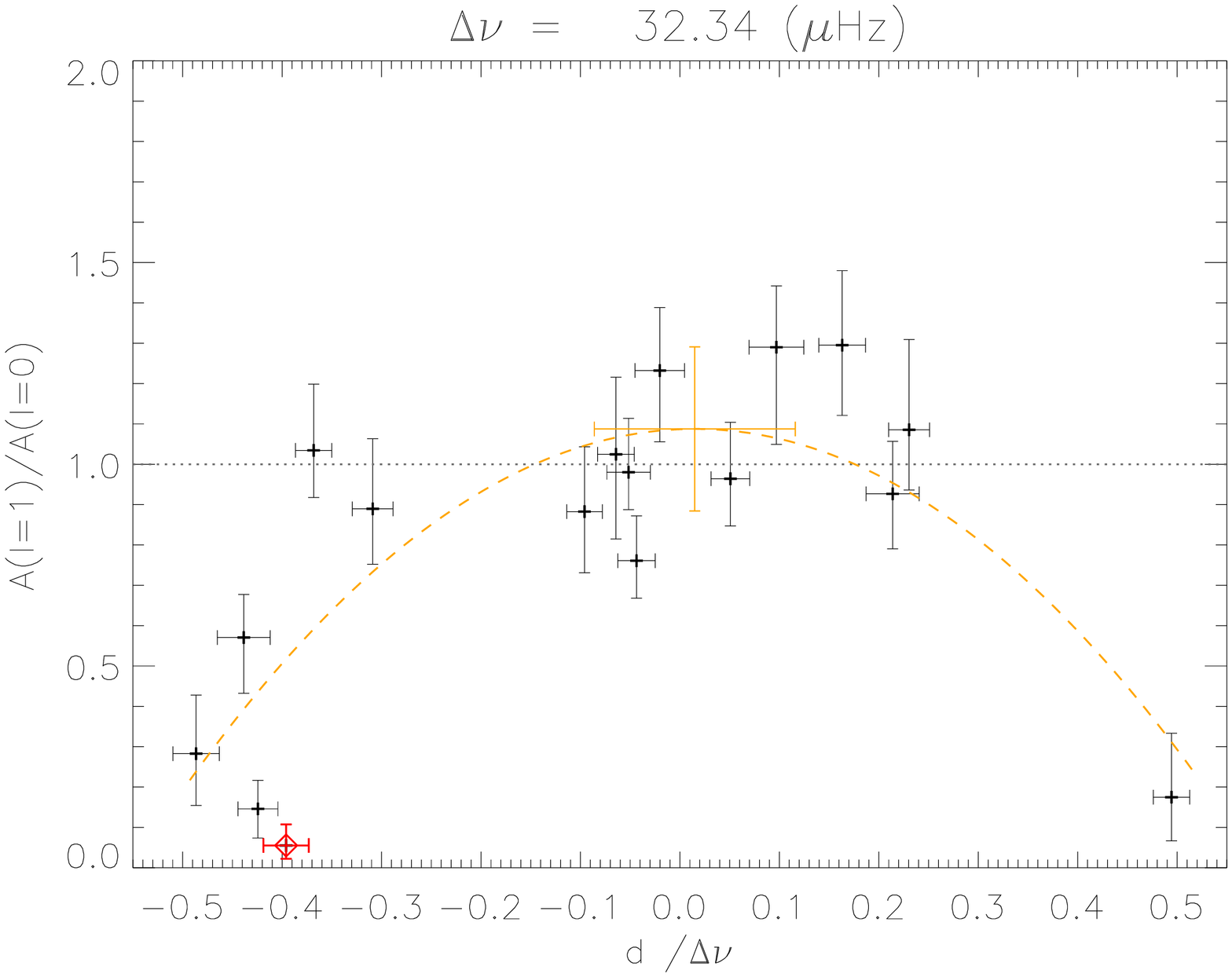} \\
\caption{$\ell=0$ (orange) and $\ell=1$ (blue) Amplitude. Vertical dashed lines indicate $\gamma$ modes positions. Vertical black dotted lines indicate $\pi$ modes positions. Red data points are not statistically significant in the power spectrum. When the $\ell=1$ mode is far from the $\pi$ mode frequencies, amplitude drops. On right, the orange dashed line is a polynomial fit performed to measure the visibilities. }
\label{fig:Amplitudes}
\end{center}
\end{figure*}

\textbf{Fig. \ref{fig:Widths} 
and \ref{fig:Amplitudes} show the Full Width at Half Maximum (FWHM) $\Gamma_\ell$ 
and the amplitudes $A_l$ (with $A_\ell=\sqrt{\pi H_\ell \Gamma_\ell}$). In each case, we show the actual value as a function of frequency (left column) and a normalized value as a function of the distance $d$, to the $\ell=1$ $\pi$-home ridge.}
The normalized values were calculated using the local, linearly interpolated $\ell=0$ values as a reference. \textbf{This allows us to compare the relative width (in $\mu$Hz) and energy (ppm) of $\ell=1$ modes for all the stars.}

We note that the width for $\ell=1$ is generally lower than for $\ell=0$ and drops at each avoided crossing. These drops are mainly an effect of the coupling between $\pi$ and $\gamma$ modes. The more gravity-like modes are narrower which can also be seen clearly in plots of FWHM versus $d/\Delta\nu$.
\textbf{A similar effect is seen more weakly in the heights of modes. 
However, when the mode FWHM approaches the spectrum resolution the height uncertainty becomes great and the stochastic excitation of the mode prevents us from providing a reliable measure}. 
\textbf{The effect of the coupling is clearest for KIC 11771760. The number of mixed modes is greater. In the $\Gamma \left( d/\Delta\nu \right)$ and $A \left( d/\Delta\nu \right)$ diagrams, the normalized quantities follow a bell shape,  centered on approximately $0$ showing that modes with mainly a p-mode character are the more intense, as expected.}

In addition to the mixing, geometrical effects and limb darkening modify the relative values of heights and amplitudes, and therefore the visibilities of each degree.
Before quantitative comparison with models, one would need to disentangle these effects. If we assume that visibilities of all $\ell=1$ modes are identical, we can estimate the visibility by measuring the relative heights or amplitudes at the home ridge, meaning when $d \approx 0$, where mixed modes behave as p modes. To do this, we fitted the measured height and amplitudes with a parabolic function. However, a stable fit was not always achievable, especially for the early subgiants. In this case, we fitted a constant around the maximum of the function $d/\Delta\nu$, by including only values that satisfies $\left| d /\Delta\nu \right| \leq  0.05$. Table \ref{tab:parameters:mes:Vis} and \ref{tab:parameters:mes:Vis2} summarize the obtained visibilities for all non-radial modes in height and amplitude. For comparison purpose, solar height visibilities \citep{Salabert2011} and for average RGB stars visibilities \citep{Mosser2012d} are also shown.

The $\ell=1$ relative heights suffer from high uncertainties. However the early subgiant (KIC 6442183 and KIC 11026764) visibilities are rather high compared to the Sun and RGB stars while for stars in the late subgiant phase (KIC 12508433 and KIC 11771760), values become comparable to RGB stars. It is clear that $\ell=2$ heights relative to $\ell=0$, are substantially higher than in the Sun or RGB for KIC 11026764 and KIC 12508433. This has been already noticed in several subgiant stars (e.g. \cite{Deheuvels2010a, Mathur2013}), although with greater uncertainties. The $\ell=3$ visibilities are consistent with solar visibilities for both subgiants and RGBs. 
According to \cite{Ballot2011} the visibility factors depend on the effective temperature and on the metallicity. However, these cannot entirely explain the discrepancy between main-sequence, subgiant and giant stars. 

\begin{deluxetable*}{cccccccccc}\tabletypesize{\footnotesize}
\tablecolumns{10}
\tablewidth{0pc}
\tablecaption{Table of computed seismic parameters for the four stars. }
\tablehead{
\colhead{KIC} & 
\colhead{$\nu_\mathrm{max}$ ($\mu$Hz)} &
\colhead{$\Delta\nu$ ($\mu$Hz)} &
\colhead{$\epsilon_p$ } &
\colhead{$\Delta{\Pi_1}$ (sec) } &  
\colhead{$\epsilon_{\gamma,1}$ } &
\colhead{$\delta\nu_{01}$ ($\mu$Hz) } &
\colhead{$\delta\nu_{02}$ ($\mu$Hz)} &
\colhead{$\delta\nu_{13}$ ($\mu$Hz) } &
}
\startdata
6442183  & $1160 \pm 4$ & $65.07 \pm 0.09$   &  $1.467 \pm 0.021$   & $325 \pm 18$ & $0.05 \pm 0.11$  &  
																																																				$4.2 \pm 0.2$   &  $5.60 \pm 0.20$  & $7.00 \pm 0.23$ \\
11026764 & $880 \pm 5$ & $50.17 \pm 0.09$   &  $1.453 \pm 0.025$   & $301 \pm 16$ & $0.59 \pm 0.20$  & 
																																																				$1.45 \pm 0.6$  &  $4.14 \pm 0.08$   & / \\
12508433 & $773 \pm 5$ & $45.18 \pm 0.05$   &  $1.407 \pm 0.008$   & $157.5 \pm 0.8 $ & $0.524 \pm 0.034$  &  
																																																					$-0.7 \pm 0.2$  &  $4.02 \pm 0.10$   & $10.91 \pm 0.10$  \\
11771760 & $522 \pm 7$ & $32.34 \pm 0.09$   &  $1.206 \pm 0.036$   & $177 \pm 5$ & $0.173  (+0.30 -0.24)$  &  
																																																					$1.28 \pm 0.38$   &  $3.00 \pm 0.09$   & / \\
\enddata \label{tab:parameters:mes}
\end{deluxetable*}

\begin{deluxetable}{cccccccc}\tabletypesize{\footnotesize}
\tablecolumns{7}
\tablewidth{0pc}
\tablecaption{Effective temperature and derived seismic radius and mass of the four stars using scaling relations. Typical uncertainties on mass and radius are a few percent.} 
\tablehead{
\colhead{KIC} & 
\colhead{$T_\mathrm{eff}$ (K)} &
\colhead{$M$ ($M_\odot$)} &
\colhead{$R$ ($R_\odot$)} 
}
\startdata
6442183  & $5740 \pm 60$    &  $\sim 0.94$  & $\sim 1.60$ \\
11026764 & $5722 \pm 57$    &  $\sim 1.14$  & $\sim 2.02$ \\
12508433 & $5257 \pm 64$    &  $\sim 1.03$  & $\sim 2.10$ \\
11771760 & $6075 \pm 70$    &  $\sim 1.51$  & $\sim 2.98$
\enddata \label{tab:parameters:Stellar}
\end{deluxetable}

\begin{deluxetable}{cccc}\tabletypesize{\footnotesize}
\tablecolumns{4}
\tablewidth{0pc}
\tablecaption{Height visibility factors for $\ell=1,2,3$ modes.}
\tablehead{
\colhead{KIC} & 
\colhead{$H_\ell=1 / H_\ell=0$ } &  
\colhead{$H_\ell=2 / H_\ell=0$ } & 
\colhead{$H_\ell=3 / H_\ell=0$ } 
}
\startdata
6442183  & $1.88  \pm 0.15$   &  $0.584 \pm 0.055$   & $0.081 \pm 0.015$   \\
11026764 & $2.04 \pm 0.21$   &  $0.773 \pm 0.081$   & $/$ \\
12508433 & $1.50 \pm 0.31$   &  $0.581 \pm 0.050$  & $0.082 \pm 0.016$   \\
11771760 & $1.26 \pm 0.10$   &  $0.776 \pm 0.088$  & $/$ \\ \hline \hline
Sun      & $1.53$ &  $0.56$  & $0.034$ \\
RGB      & $1.35$            &  $0.64$             & $0.071$ 
\enddata \label{tab:parameters:mes:Vis}
\end{deluxetable}

\begin{deluxetable}{cccc}\tabletypesize{\footnotesize}
\tablecolumns{4}
\tablewidth{0pc}
\tablecaption{Amplitude visibility factors for $\ell=1,2,3$ modes. }
\tablehead{
\colhead{KIC} & 
\colhead{$A_\ell=1  / A_\ell=0$ } &  
\colhead{$A_\ell=2 / A_\ell=0$ } & 
\colhead{$A_\ell=3 / A_\ell=0$ }
}
\startdata
6442183  & $1.27 \pm 0.15$   &  $0.751  \pm 0.061$   & $0.285 \pm 0.032$   \\
11026764 & $1.37 \pm 0.12$   &  $0.875 \pm 0.082$   & $/$ \\
12508433 & $1.18 \pm 0.09$   &  $0.768 \pm 0.063$  & $0.272 \pm 0.034$   \\
11771760 & $1.09 \pm 0.14$   &  $0.809 \pm 0.075$  & $/$
\enddata \label{tab:parameters:mes:Vis2}
\end{deluxetable}

\begin{deluxetable*}{cccccccccc}\tabletypesize{\footnotesize}
\tablecolumns{10}
\tablewidth{0pc}
\tablecaption{Observed mode parameters and their uncertainties  for KIC 6442183.}
\tablehead{
\multicolumn{10}{c}{KIC 6442183 - Mode range: [700 - 1500] $\mu$Hz} \\ \hline
\colhead{$\ell$} & 
\colhead{$\nu$ ($\mu$Hz)} &  
\colhead{$err_{+}(\nu)$ } & 
\colhead{$err_{-}(\nu)$ } & 
\colhead{$\Gamma$ ($\mu$Hz)} & 
\colhead{$err_{-}(\Gamma)$ } & 
\colhead{$err_{+}(\Gamma)$ } & 
\colhead{A (ppm)} & 
\colhead{$err_{-}(A)$ } & 
\colhead{$err_{+}(A)$ }
}
\startdata
$       0$ & $    743.47$ & $    0.11$ & $    0.11$ & $    0.64$ & $    0.14$ & $    0.13$ & $    2.59$ & $    0.48$ & $    0.46$ \\
$       0$ & $    809.55$ & $    0.23$ & $    0.16$ & $    0.88$ & $    0.22$ & $    0.24$ & $    2.51$ & $    0.45$ & $    0.45$ \\
$       0$ & $    873.85$ & $    0.10$ & $    0.10$ & $    0.83$ & $    0.14$ & $    0.17$ & $    4.06$ & $    0.39$ & $    0.39$ \\
$       0$ & $    937.14$ & $    0.10$ & $    0.10$ & $    0.82$ & $    0.11$ & $    0.14$ & $    4.73$ & $    0.37$ & $    0.39$ \\
$       0$ & $   1001.04$ & $    0.09$ & $    0.09$ & $    0.74$ & $    0.09$ & $    0.11$ & $    6.06$ & $    0.49$ & $    0.52$ \\
$       0$ & $   1065.63$ & $    0.09$ & $    0.09$ & $    0.79$ & $    0.10$ & $    0.11$ & $    6.46$ & $    0.42$ & $    0.45$ \\
$       0$ & $   1130.49$ & $    0.06$ & $    0.06$ & $    0.69$ & $    0.08$ & $    0.09$ & $    9.61$ & $    0.53$ & $    0.55$ \\
$       0$ & $   1195.41$ & $    0.06$ & $    0.06$ & $    0.66$ & $    0.08$ & $    0.08$ & $    9.87$ & $    0.55$ & $    0.60$ \\
$       0$ & $   1260.23$ & $    0.08$ & $    0.09$ & $    1.13$ & $    0.13$ & $    0.14$ & $    7.85$ & $    0.41$ & $    0.43$ \\
$       0$ & $   1325.63$ & $    0.14$ & $    0.13$ & $    1.46$ & $    0.19$ & $    0.21$ & $    6.58$ & $    0.39$ & $    0.40$ \\
$       0$ & $   1391.92$ & $    0.22$ & $    0.22$ & $    2.07$ & $    0.28$ & $    0.37$ & $    4.76$ & $    0.33$ & $    0.33$ \\
$       0$ & $   1458.12$ & $    0.53$ & $    0.47$ & $    3.37$ & $    0.49$ & $    0.58$ & $    3.31$ & $    0.29$ & $    0.29$ \\
$       1$ & $    745.74$ & $    5.61$ & $    0.24$ & $    0.25$ & $    0.16$ & $    0.29$ & $    1.74$ & $    0.62$ & $    0.62$ \\
$       1$ & $    783.04$ & $    0.30$ & $    0.31$ & $    0.64$ & $    0.22$ & $    0.20$ & $    3.18$ & $    0.54$ & $    0.53$ \\
$       1$ & $    840.74$ & $    0.09$ & $    0.09$ & $    0.62$ & $    0.18$ & $    0.22$ & $    4.26$ & $    0.49$ & $    0.49$ \\
$       1$ & $    901.28$ & $    0.09$ & $    0.10$ & $    0.70$ & $    0.16$ & $    0.17$ & $    4.73$ & $    0.45$ & $    0.48$ \\
$       1$ & $    960.29$ & $    0.10$ & $    0.10$ & $    0.82$ & $    0.15$ & $    0.15$ & $    5.69$ & $    0.48$ & $    0.53$ \\
$       1$ & $   1002.93$ & $    0.06$ & $    0.05$ & $    0.26$ & $    0.09$ & $    0.18$ & $    6.17$ & $    0.80$ & $    0.89$ \\
$       1$ & $   1037.59$ & $    0.07$ & $    0.07$ & $    0.64$ & $    0.13$ & $    0.13$ & $    8.07$ & $    0.60$ & $    0.67$ \\
$       1$ & $   1096.78$ & $    0.05$ & $    0.05$ & $    0.51$ & $    0.10$ & $    0.12$ & $    9.98$ & $    0.78$ & $    0.89$ \\
$       1$ & $   1159.92$ & $    0.05$ & $    0.05$ & $    0.55$ & $    0.10$ & $    0.10$ & $   13.15$ & $    0.90$ & $    1.17$ \\
$       1$ & $   1224.10$ & $    0.07$ & $    0.06$ & $    0.85$ & $    0.12$ & $    0.12$ & $   11.32$ & $    0.68$ & $    0.73$ \\
$       1$ & $   1288.63$ & $    0.09$ & $    0.09$ & $    1.23$ & $    0.17$ & $    0.16$ & $    8.99$ & $    0.51$ & $    0.55$ \\
$       1$ & $   1353.98$ & $    0.12$ & $    0.12$ & $    1.71$ & $    0.22$ & $    0.23$ & $    7.78$ & $    0.40$ & $    0.45$ \\
$       1$ & $   1419.44$ & $    0.14$ & $    0.13$ & $    1.67$ & $    0.30$ & $    0.34$ & $    5.82$ & $    0.37$ & $    0.39$ \\
$       2$ & $    737.14$ & $    2.54$ & $    1.11$ & $    0.58$ & $    0.13$ & $    0.15$ & $    1.92$ & $    0.39$ & $    0.40$ \\
$       2$ & $    801.89$ & $    0.32$ & $    0.70$ & $    0.90$ & $    0.25$ & $    0.28$ & $    2.01$ & $    0.36$ & $    0.35$ \\
$       2$ & $    867.84$ & $    0.20$ & $    0.19$ & $    0.78$ & $    0.18$ & $    0.22$ & $    2.91$ & $    0.33$ & $    0.33$ \\
$       2$ & $    931.41$ & $    0.14$ & $    0.15$ & $    0.82$ & $    0.15$ & $    0.18$ & $    3.58$ & $    0.31$ & $    0.34$ \\
$       2$ & $    996.73$ & $    0.10$ & $    0.10$ & $    0.52$ & $    0.10$ & $    0.12$ & $    3.79$ & $    0.34$ & $    0.35$ \\
$       2$ & $   1059.92$ & $    0.08$ & $    0.07$ & $    0.71$ & $    0.11$ & $    0.13$ & $    4.71$ & $    0.32$ & $    0.33$ \\
$       2$ & $   1124.66$ & $    0.06$ & $    0.06$ & $    0.63$ & $    0.08$ & $    0.09$ & $    6.82$ & $    0.40$ & $    0.42$ \\
$       2$ & $   1189.77$ & $    0.07$ & $    0.07$ & $    0.72$ & $    0.09$ & $    0.11$ & $    7.88$ & $    0.41$ & $    0.44$ \\
$       2$ & $   1254.54$ & $    0.11$ & $    0.11$ & $    1.20$ & $    0.15$ & $    0.16$ & $    6.67$ & $    0.36$ & $    0.35$ \\
$       2$ & $   1321.35$ & $    0.14$ & $    0.14$ & $    1.25$ & $    0.21$ & $    0.24$ & $    4.79$ & $    0.33$ & $    0.33$ \\
$       2$ & $   1386.89$ & $    0.21$ & $    0.20$ & $    2.30$ & $    0.36$ & $    0.43$ & $    4.08$ & $    0.26$ & $    0.28$ \\
$       2$ & $   1453.09$ & $    0.38$ & $    0.38$ & $    3.08$ & $    0.55$ & $    0.65$ & $    2.63$ & $    0.22$ & $    0.22$ \\
$       3$ & $   1085.64$ & $    1.01$ & $    0.32$ & $    0.76$ & $    0.08$ & $    0.09$ & $    2.19$ & $    0.20$ & $    0.22$ \\
$       3$ & $   1150.25$ & $    0.17$ & $    0.17$ & $    0.76$ & $    0.11$ & $    0.12$ & $    2.92$ & $    0.26$ & $    0.28$ \\
$       3$ & $   1215.67$ & $    0.18$ & $    0.15$ & $    0.66$ & $    0.15$ & $    0.16$ & $    2.51$ & $    0.28$ & $    0.29$ \\
$       3$ & $   1280.75$ & $    0.39$ & $    0.35$ & $    1.42$ & $    0.28$ & $    0.31$ & $    2.34$ & $    0.27$ & $    0.27$ \\
$       3$ & $   1347.01$ & $    0.37$ & $    0.41$ & $    1.41$ & $    0.32$ & $    0.37$ & $    1.65$ & $    0.22$ & $    0.22$ \\
$       3$ & $   1413.29$ & $    0.77$ & $    0.71$ & $    2.50$ & $    0.36$ & $    0.43$ & $    1.31$ & $    0.15$ & $    0.16$ \\
\enddata \label{tab:parameters:mes:all:KIC6442183}
\end{deluxetable*}

\begin{deluxetable*}{cccccccccc}\tabletypesize{\footnotesize}
\tablecolumns{10}
\tablewidth{0pc}
\tablecaption{Observed mode parrameters and their uncertainties  for KIC 11026764.}
\tablehead{
\multicolumn{10}{c}{KIC 11026764 - Mode range: [550 - 1100] $\mu$Hz} \\ \hline
\colhead{$\ell$} & 
\colhead{$\nu$ ($\mu$Hz)} &  
\colhead{$err_{+}(\nu)$ } & 
\colhead{$err_{-}(\nu)$ } & 
\colhead{$\Gamma$ ($\mu$Hz)} & 
\colhead{$err_{-}(\Gamma)$ } & 
\colhead{$err_{+}(\Gamma)$ } & 
\colhead{A (ppm)} & 
\colhead{$err_{-}(A)$ } & 
\colhead{$err_{+}(A)$ }
}
\startdata
$       0$ & $    574.68$ & $    0.11$ & $    0.16$ & $    0.39$ & $    0.14$ & $    0.23$ & $    2.98$ & $    0.47$ & $    0.52$ \\
$       0$ & $    625.01$ & $    0.10$ & $    0.11$ & $    0.40$ & $    0.10$ & $    0.13$ & $    3.92$ & $    0.43$ & $    0.47$ \\
$       0$ & $    674.53$ & $    0.09$ & $    0.10$ & $    0.52$ & $    0.14$ & $    0.24$ & $    3.76$ & $    0.44$ & $    0.47$ \\
$       0$ & $    723.48$ & $    0.15$ & $    0.14$ & $    0.99$ & $    0.20$ & $    0.25$ & $    5.11$ & $    0.50$ & $    0.49$ \\
$       0$ & $    773.00$ & $    0.11$ & $    0.09$ & $    0.85$ & $    0.16$ & $    0.19$ & $    4.59$ & $    0.37$ & $    0.42$ \\
$       0$ & $    823.21$ & $    0.07$ & $    0.07$ & $    0.76$ & $    0.11$ & $    0.13$ & $    7.93$ & $    0.51$ & $    0.52$ \\
$       0$ & $    873.75$ & $    0.07$ & $    0.07$ & $    0.78$ & $    0.10$ & $    0.12$ & $    8.64$ & $    0.53$ & $    0.56$ \\
$       0$ & $    924.06$ & $    0.10$ & $    0.10$ & $    0.90$ & $    0.14$ & $    0.16$ & $    8.27$ & $    0.57$ & $    0.55$ \\
$       0$ & $    974.56$ & $    0.15$ & $    0.14$ & $    1.22$ & $    0.17$ & $    0.21$ & $    6.99$ & $    0.46$ & $    0.48$ \\
$       0$ & $   1025.49$ & $    0.17$ & $    0.17$ & $    1.69$ & $    0.26$ & $    0.34$ & $    5.61$ & $    0.40$ & $    0.41$ \\
$       0$ & $   1076.10$ & $    0.52$ & $    0.42$ & $    2.92$ & $    0.55$ & $    0.75$ & $    3.66$ & $    0.36$ & $    0.37$ \\
$       1$ & $    694.12$ & $    0.06$ & $    0.06$ & $    0.37$ & $    0.13$ & $    0.16$ & $    5.14$ & $    0.58$ & $    0.72$ \\
$       1$ & $    723.91$ & $    0.33$ & $    2.47$ & $    0.03$ & $    0.02$ & $    0.16$ & $    2.15$ & $    1.16$ & $    1.44$ \\
$       1$ & $    755.10$ & $    0.10$ & $    0.11$ & $    0.83$ & $    0.17$ & $    0.16$ & $    6.45$ & $    0.53$ & $    0.63$ \\
$       1$ & $    799.62$ & $    0.08$ & $    0.08$ & $    0.78$ & $    0.14$ & $    0.14$ & $    8.91$ & $    0.62$ & $    0.73$ \\
$       1$ & $    847.39$ & $    0.07$ & $    0.07$ & $    0.73$ & $    0.17$ & $    0.14$ & $    9.85$ & $    0.74$ & $    1.10$ \\
$       1$ & $    894.03$ & $    0.06$ & $    0.06$ & $    0.67$ & $    0.13$ & $    0.14$ & $   11.56$ & $    0.77$ & $    0.92$ \\
$       1$ & $    925.83$ & $    0.06$ & $    0.06$ & $    0.39$ & $    0.12$ & $    0.12$ & $    7.99$ & $    0.81$ & $    0.98$ \\
$       1$ & $    954.09$ & $    0.08$ & $    0.08$ & $    0.96$ & $    0.14$ & $    0.15$ & $   10.67$ & $    0.64$ & $    0.70$ \\
$       1$ & $   1000.41$ & $    0.08$ & $    0.08$ & $    1.02$ & $    0.18$ & $    0.21$ & $    8.96$ & $    0.60$ & $    0.67$ \\
$       1$ & $   1050.11$ & $    0.14$ & $    0.14$ & $    1.70$ & $    0.33$ & $    0.37$ & $    6.20$ & $    0.41$ & $    0.46$ \\
$       2$ & $    570.32$ & $    0.18$ & $    0.14$ & $    0.37$ & $    0.15$ & $    0.23$ & $    2.44$ & $    0.45$ & $    0.50$ \\
$       2$ & $    620.64$ & $    0.11$ & $    0.09$ & $    0.42$ & $    0.12$ & $    0.15$ & $    3.48$ & $    0.40$ & $    0.43$ \\
$       2$ & $    669.49$ & $    0.15$ & $    0.13$ & $    0.41$ & $    0.15$ & $    0.23$ & $    3.03$ & $    0.41$ & $    0.41$ \\
$       2$ & $    718.92$ & $    0.22$ & $    0.22$ & $    1.10$ & $    0.24$ & $    0.30$ & $    4.76$ & $    0.42$ & $    0.42$ \\
$       2$ & $    765.85$ & $    0.14$ & $    0.21$ & $    0.36$ & $    0.16$ & $    0.21$ & $    2.67$ & $    0.65$ & $    0.59$ \\
$       2$ & $    771.41$ & $    0.26$ & $    0.27$ & $    0.83$ & $    0.18$ & $    0.24$ & $    4.04$ & $    0.36$ & $    0.38$ \\
$       2$ & $    818.67$ & $    0.10$ & $    0.10$ & $    0.79$ & $    0.13$ & $    0.16$ & $    6.93$ & $    0.43$ & $    0.47$ \\
$       2$ & $    868.89$ & $    0.09$ & $    0.10$ & $    0.87$ & $    0.12$ & $    0.13$ & $    7.97$ & $    0.45$ & $    0.49$ \\
$       2$ & $    919.89$ & $    0.10$ & $    0.10$ & $    1.05$ & $    0.15$ & $    0.18$ & $    7.98$ & $    0.48$ & $    0.49$ \\
$       2$ & $    970.15$ & $    0.10$ & $    0.11$ & $    1.16$ & $    0.17$ & $    0.20$ & $    6.26$ & $    0.38$ & $    0.42$ \\
$       2$ & $   1020.41$ & $    0.18$ & $    0.20$ & $    1.50$ & $    0.30$ & $    0.32$ & $    4.89$ & $    0.36$ & $    0.38$ \\
$       2$ & $   1072.09$ & $    0.45$ & $    0.50$ & $    2.74$ & $    0.56$ & $    0.69$ & $    3.48$ & $    0.32$ & $    0.32$ \\
\enddata \label{tab:parameters:mes:all:KIC11026764}
\end{deluxetable*}

\begin{deluxetable*}{cccccccccc}\tabletypesize{\footnotesize}
\tablecolumns{10}
\tablewidth{0pc}
\tablecaption{Observed mode parrameters and their uncertainties  for KIC 12508433. The mode with an asterisk is non-statistically significant.}
\tablehead{
\multicolumn{10}{c}{KIC 12508433 - Mode range: [590 - 1050] $\mu$Hz} \\ \hline
\colhead{$\ell$} & 
\colhead{$\nu$ ($\mu$Hz)} &  
\colhead{$err_{+}(\nu)$ } & 
\colhead{$err_{-}(\nu)$ } & 
\colhead{$\Gamma$ ($\mu$Hz) } & 
\colhead{$err_{-}(\Gamma)$ } & 
\colhead{$err_{+}(\Gamma)$ } & 
\colhead{A (ppm)} & 
\colhead{$err_{-}(A)$ } & 
\colhead{$err_{+}(A)$ }
}
\startdata
$       0$ & $    606.10$ & $    0.05$ & $    0.06$ & $    0.41$ & $    0.09$ & $    0.12$ & $    4.53$ & $    0.45$ & $    0.46$ \\
$       0$ & $    649.96$ & $    0.06$ & $    0.06$ & $    0.47$ & $    0.08$ & $    0.10$ & $    6.31$ & $    0.56$ & $    0.57$ \\
$       0$ & $    694.41$ & $    0.05$ & $    0.05$ & $    0.45$ & $    0.05$ & $    0.06$ & $    7.89$ & $    0.61$ & $    0.64$ \\
$       0$ & $    739.42$ & $    0.05$ & $    0.05$ & $    0.44$ & $    0.05$ & $    0.06$ & $   11.19$ & $    0.82$ & $    0.90$ \\
$       0$ & $    784.16$ & $    0.05$ & $    0.05$ & $    0.42$ & $    0.06$ & $    0.07$ & $   13.00$ & $    0.83$ & $    0.91$ \\
$       0$ & $    829.08$ & $    0.06$ & $    0.06$ & $    0.54$ & $    0.07$ & $    0.08$ & $   10.42$ & $    0.68$ & $    0.72$ \\
$       0$ & $    874.25$ & $    0.07$ & $    0.07$ & $    0.69$ & $    0.08$ & $    0.10$ & $    9.20$ & $    0.60$ & $    0.65$ \\
$       0$ & $    919.82$ & $    0.15$ & $    0.15$ & $    1.23$ & $    0.16$ & $    0.17$ & $    5.68$ & $    0.45$ & $    0.49$ \\
$       0$ & $    965.75$ & $    0.21$ & $    0.23$ & $    1.58$ & $    0.38$ & $    0.44$ & $    4.02$ & $    0.45$ & $    0.45$ \\
$       0$ & $   1012.57$ & $    0.22$ & $    0.13$ & $    1.97$ & $    0.51$ & $    0.63$ & $    2.91$ & $    0.46$ & $    0.45$ \\
$       1$ & $    627.14$ & $    0.06$ & $    0.06$ & $    0.38$ & $    0.08$ & $    0.09$ & $    5.31$ & $    0.58$ & $    0.61$ \\
$       1$ & $    657.25$ & $    0.07$ & $    0.07$ & $    0.26$ & $    0.05$ & $    0.08$ & $    7.51$ & $    0.78$ & $    0.72$ \\
$       1$ & $    679.15$ & $    0.05$ & $    0.06$ & $    0.37$ & $    0.09$ & $    0.09$ & $    7.01$ & $    0.72$ & $    0.78$ \\
$       1$ & $    713.54$ & $    0.05$ & $    0.05$ & $    0.35$ & $    0.07$ & $    0.10$ & $    8.84$ & $    0.83$ & $    0.89$ \\
$       1$ & $    740.66$ & $    0.07$ & $    0.06$ & $    0.25$ & $    0.11$ & $    0.11$ & $    8.42$ & $    1.20$ & $    1.49$ \\
$       1$ & $    768.29$ & $    0.04$ & $    0.04$ & $    0.32$ & $    0.04$ & $    0.06$ & $   13.98$ & $    0.97$ & $    0.90$ \\
$       1$ & $    806.14$ & $    0.04$ & $    0.04$ & $    0.32$ & $    0.04$ & $    0.05$ & $   15.21$ & $    0.92$ & $    0.92$ \\
$       1$ & $    836.05$ & $    0.04$ & $    0.04$ & $    0.31$ & $    0.05$ & $    0.07$ & $   11.46$ & $    0.98$ & $    0.92$ \\
$       1$ & $    860.47$ & $    0.06$ & $    0.05$ & $    0.51$ & $    0.09$ & $    0.12$ & $    9.04$ & $    0.77$ & $    0.81$ \\
$       1$ & $    899.99$ & $    0.06$ & $    0.07$ & $    0.59$ & $    0.15$ & $    0.14$ & $    9.14$ & $    0.80$ & $    1.11$ \\
$       1$ & $    942.31$ & $    0.14$ & $    0.14$ & $    1.27$ & $    0.25$ & $    0.31$ & $    6.02$ & $    0.54$ & $    0.55$ \\
$       1$* & $    974.11$ & $    0.14$ & $    0.15$ & $    0.01$ & $    0.01$ & $    0.01$ & $    0.86$ & $    0.41$ & $    0.47$ \\
$       1$ & $   1000.05$ & $    0.31$ & $    0.33$ & $    1.70$ & $    0.57$ & $    0.64$ & $    4.37$ & $    0.54$ & $    0.54$ \\
$       2$ & $    602.73$ & $    0.18$ & $    0.19$ & $    0.40$ & $    0.10$ & $    0.14$ & $    3.30$ & $    0.42$ & $    0.46$ \\
$       2$ & $    646.12$ & $    0.14$ & $    0.12$ & $    0.49$ & $    0.10$ & $    0.12$ & $    4.86$ & $    0.46$ & $    0.47$ \\
$       2$ & $    690.61$ & $    0.14$ & $    0.14$ & $    0.44$ & $    0.08$ & $    0.09$ & $    5.88$ & $    0.53$ & $    0.56$ \\
$       2$ & $    735.95$ & $    0.06$ & $    0.06$ & $    0.47$ & $    0.06$ & $    0.07$ & $    8.65$ & $    0.57$ & $    0.58$ \\
$       2$ & $    779.60$ & $    0.05$ & $    0.05$ & $    0.43$ & $    0.06$ & $    0.08$ & $    9.90$ & $    0.62$ & $    0.64$ \\
$       2$ & $    825.20$ & $    0.07$ & $    0.08$ & $    0.58$ & $    0.08$ & $    0.09$ & $    8.64$ & $    0.53$ & $    0.57$ \\
$       2$ & $    870.92$ & $    0.08$ & $    0.09$ & $    0.68$ & $    0.10$ & $    0.11$ & $    7.12$ & $    0.43$ & $    0.48$ \\
$       2$ & $    915.61$ & $    0.18$ & $    0.19$ & $    1.34$ & $    0.21$ & $    0.24$ & $    5.27$ & $    0.40$ & $    0.42$ \\
$       2$ & $    961.35$ & $    0.28$ & $    0.56$ & $    1.40$ & $    0.44$ & $    0.50$ & $    3.08$ & $    0.35$ & $    0.35$ \\
$       2$ & $   1006.98$ & $    0.55$ & $    0.57$ & $    1.92$ & $    0.52$ & $    0.70$ & $    2.39$ & $    0.34$ & $    0.34$ \\
$       3$ & $    752.04$ & $    0.23$ & $    0.27$ & $    0.44$ & $    0.05$ & $    0.06$ & $    3.23$ & $    0.37$ & $    0.39$ \\
$       3$ & $    797.26$ & $    0.13$ & $    0.16$ & $    0.44$ & $    0.08$ & $    0.09$ & $    3.33$ & $    0.38$ & $    0.40$ \\
$       3$ & $    842.53$ & $    0.18$ & $    0.24$ & $    0.54$ & $    0.10$ & $    0.11$ & $    2.66$ & $    0.33$ & $    0.36$ \\
$       3$ & $    887.44$ & $    0.30$ & $    0.38$ & $    0.90$ & $    0.13$ & $    0.13$ & $    2.50$ & $    0.30$ & $    0.31$ \\
\enddata \label{tab:parameters:mes:all:KIC12508433}
\end{deluxetable*}

\begin{deluxetable*}{cccccccccc}\tabletypesize{\footnotesize}
\tablecolumns{10}
\tablewidth{0pc}
\tablecaption{Observed mode parameters and their uncertainties for KIC 11771760. The mode with an asterisk is non-statistically significant.}
\tablehead{
\multicolumn{10}{c}{KIC 11771760 - Mode range: [350 - 700] $\mu$Hz} \\ \hline
\colhead{$\ell$} & 
\colhead{$\nu$ ($\mu$Hz)} &  
\colhead{$err_{+}(\nu)$ } & 
\colhead{$err_{-}(\nu)$ } & 
\colhead{$\Gamma$ ($\mu$Hz)} & 
\colhead{$err_{-}(\Gamma)$ } & 
\colhead{$err_{+}(\Gamma)$ } & 
\colhead{A (ppm)} & 
\colhead{$err_{-}(A)$ } & 
\colhead{$err_{+}(A)$ }
}
\startdata
$       0$ & $    361.26$ & $    0.08$ & $    0.12$ & $    1.02$ & $    0.23$ & $    0.24$ & $    8.25$ & $    1.07$ & $    1.06$ \\
$       0$ & $    394.35$ & $    0.18$ & $    0.18$ & $    1.13$ & $    0.24$ & $    0.31$ & $    5.49$ & $    0.66$ & $    0.58$ \\
$       0$ & $    426.66$ & $    0.10$ & $    0.11$ & $    0.81$ & $    0.10$ & $    0.11$ & $   10.32$ & $    0.67$ & $    0.62$ \\
$       0$ & $    457.27$ & $    0.11$ & $    0.10$ & $    1.12$ & $    0.11$ & $    0.11$ & $   14.38$ & $    0.68$ & $    0.80$ \\
$       0$ & $    488.96$ & $    0.17$ & $    0.16$ & $    1.01$ & $    0.17$ & $    0.23$ & $    8.39$ & $    0.75$ & $    1.08$ \\
$       0$ & $    521.59$ & $    0.13$ & $    0.16$ & $    1.06$ & $    0.10$ & $    0.12$ & $   12.20$ & $    0.61$ & $    0.55$ \\
$       0$ & $    554.69$ & $    0.10$ & $    0.08$ & $    1.05$ & $    0.11$ & $    0.13$ & $   14.95$ & $    0.70$ & $    0.79$ \\
$       0$ & $    586.88$ & $    0.18$ & $    0.40$ & $    2.19$ & $    0.26$ & $    0.18$ & $    7.78$ & $    0.67$ & $    0.79$ \\
$       0$ & $    619.55$ & $    0.28$ & $    0.17$ & $    1.45$ & $    0.18$ & $    0.18$ & $   10.38$ & $    0.85$ & $    0.71$ \\
$       0$ & $    651.92$ & $    0.11$ & $    0.47$ & $    1.20$ & $    0.13$ & $    0.13$ & $    8.39$ & $    0.54$ & $    0.82$ \\
$       1$ & $    364.83$ & $    0.24$ & $    0.21$ & $    1.11$ & $    0.24$ & $    0.37$ & $    8.22$ & $    0.68$ & $    0.95$ \\
$       1$ & $    374.66$ & $    0.27$ & $    0.21$ & $    0.87$ & $    0.41$ & $    0.45$ & $    7.31$ & $    1.32$ & $    1.09$ \\
$       1$ & $    393.19$ & $    0.50$ & $    0.50$ & $    0.37$ & $    0.25$ & $    0.29$ & $    1.58$ & $    0.68$ & $    0.79$ \\
$       1$ & $    405.82$ & $    0.12$ & $    0.11$ & $    0.74$ & $    0.29$ & $    0.26$ & $    6.36$ & $    0.88$ & $    0.71$ \\
$       1$ & $    424.91$ & $    0.20$ & $    0.14$ & $    0.40$ & $    0.28$ & $    0.20$ & $    1.76$ & $    1.08$ & $    1.57$ \\
$       1$ & $    439.69$ & $    0.14$ & $    0.14$ & $    0.82$ & $    0.17$ & $    0.17$ & $    9.17$ & $    0.95$ & $    0.88$ \\
$       1$ & $    459.56$ & $    0.21$ & $    0.16$ & $    0.64$ & $    0.52$ & $    0.20$ & $    2.04$ & $    1.00$ & $    0.98$ \\
$       1$ & $    474.92$ & $    0.10$ & $    0.09$ & $    0.73$ & $    0.14$ & $    0.15$ & $   10.65$ & $    0.93$ & $    1.13$ \\
$       1$ & $    495.47$ & $    0.08$ & $    0.09$ & $    0.55$ & $    0.13$ & $    0.13$ & $    8.14$ & $    0.93$ & $    0.82$ \\
$       1$ & $    512.92$ & $    0.10$ & $    0.10$ & $    0.88$ & $    0.19$ & $    0.18$ & $   12.14$ & $    1.03$ & $    1.22$ \\
$       1$ & $    535.97$ & $    0.12$ & $    0.11$ & $    1.07$ & $    0.12$ & $    0.10$ & $   13.13$ & $    0.69$ & $    0.68$ \\
$       1$* & $    557.02$ & $    0.03$ & $    0.05$ & $    0.20$ & $    0.14$ & $    0.20$ & $    0.80$ & $    0.47$ & $    0.75$ \\
$       1$ & $    575.10$ & $    0.20$ & $    0.18$ & $    1.92$ & $    0.14$ & $    0.17$ & $   13.47$ & $    1.14$ & $    1.37$ \\
$       1$ & $    601.35$ & $    0.21$ & $    0.20$ & $    1.75$ & $    0.20$ & $    0.19$ & $   11.00$ & $    0.92$ & $    0.74$ \\
$       1$ & $    620.00$ & $    0.23$ & $    0.26$ & $    0.87$ & $    0.16$ & $    0.24$ & $    5.90$ & $    1.37$ & $    1.01$ \\
$       1$ & $    641.11$ & $    0.26$ & $    0.25$ & $    1.30$ & $    0.22$ & $    0.20$ & $    8.39$ & $    0.97$ & $    0.96$ \\
$       1$ & $    669.51$ & $    0.19$ & $    0.20$ & $    1.07$ & $    0.21$ & $    0.14$ & $    9.43$ & $    1.21$ & $    0.70$ \\
$       2$ & $    359.35$ & $    0.13$ & $    0.15$ & $    0.86$ & $    0.15$ & $    0.18$ & $    6.15$ & $    0.82$ & $    1.11$ \\
$       2$ & $    390.57$ & $    0.18$ & $    0.24$ & $    1.16$ & $    0.31$ & $    0.27$ & $    4.84$ & $    0.82$ & $    0.54$ \\
$       2$ & $    422.54$ & $    0.12$ & $    0.13$ & $    0.81$ & $    0.10$ & $    0.11$ & $    7.95$ & $    0.65$ & $    0.69$ \\
$       2$ & $    454.72$ & $    0.21$ & $    0.22$ & $    1.00$ & $    0.12$ & $    0.13$ & $   10.85$ & $    0.84$ & $    1.07$ \\
$       2$ & $    486.64$ & $    0.17$ & $    0.16$ & $    1.09$ & $    0.22$ & $    0.21$ & $    7.16$ & $    0.66$ & $    1.42$ \\
$       2$ & $    518.58$ & $    0.12$ & $    0.12$ & $    1.09$ & $    0.14$ & $    0.13$ & $    9.85$ & $    1.16$ & $    0.68$ \\
$       2$ & $    551.08$ & $    0.13$ & $    0.14$ & $    0.80$ & $    0.13$ & $    0.14$ & $   10.34$ & $    0.89$ & $    0.96$ \\
$       2$ & $    584.11$ & $    0.45$ & $    0.59$ & $    2.44$ & $    0.21$ & $    0.25$ & $    8.50$ & $    0.90$ & $    0.65$ \\
$       2$ & $    616.71$ & $    0.53$ & $    0.38$ & $    1.51$ & $    0.25$ & $    0.27$ & $    8.24$ & $    0.59$ & $    0.62$ \\
$       2$ & $    648.92$ & $    0.17$ & $    0.18$ & $    1.26$ & $    0.17$ & $    0.18$ & $    7.06$ & $    0.47$ & $    0.62$ \\
\enddata \label{tab:parameters:mes:all:KIC11771760}
\end{deluxetable*}

\section{Discussion} \label{sec:6}
\
\textbf{Precise measures of all the properties of the oscillations in early subgiants (\emph{i.e.} when $N_\gamma \ll N_\pi$) are achieved with the coupled oscillator model without any problem. Inacurate reconstruction of the $\pi$-mode and $\gamma$-mode spectrum (artifacts) may yet happen in late giants with the current methodology. However, because the more evolved stars have a very dense $\gamma$-mode spectra ($N_\gamma \gg N_\pi$) a rich mixed mode spectrum is observed that might provide more stringent constraints on the internal structure for giants than for subgiants. Further work might therefore focus on late giants if one wants to, for example, apply inversion methods to constrain the internal stellar structure using the widths and amplitudes of mixed modes.}

In this study, we restricted our analysis to the simplest subgiants, without visible rotation, with a low gravity-mode density and at high signal to noise. However, in many cases, observed subgiants and giants stars present a clear rotational signature and often have a radial differential rotation, implying important variations of splitting from one ridge to another. Modes width could then be overestimated in the case of a non-negligible splitting. With stellar cores, rotating faster than the surface (\emph{e.g.} \citealt{Deheuvels2012}), the bias will be more significant for modes far from the p modes home ridge.

A more sophisticated analysis, which has still to be developed, would be necessary in order to measure precisely height, widths and amplitudes, but also rotational (and possible magnetic) splittings of all modes. At the moment, global approaches that use the coupled oscillator model require MCMC fits to provide reliable results and are complex. This study is at the edge of what currently could be possibly done. The main obstacle is the computational time, which increases quickly with the number of degrees of freedom of the fit. To overcome this difficulty, a local fitting approach, focusing on the stars with long-continuous time-series and with the highest signal-to-noise is probably more suited.

In this paper, we also quantified for the first time the effect of the coupling between pressure and gravity modes on widths, heights and amplitudes. Qualitatively our results are in agreement with \cite{Dupret2009}, however a quantitative comparison would certainly give a new insight on stellar interiors. 

Interestingly, KIC 11026764 and KIC 12508433 have clear bumped $\ell=2$ modes. As for $\ell=1$ modes, this mode bumping is a signature of their mixed character. However, $\ell=2$ mixed modes have a weaker coupling than in $\ell=1$ modes and therefore avoided crossings are not easy to detect in the \'echelle diagram. \textbf{For KIC 11026764, the two mixed modes are measurable. But for KIC 12058433 only a bumping on one of them is visible, the second mixed mode being either too faint or not resolved.} 

A combined measure of $\ell=1$ and $\ell=2$ mixed modes allows us to better constrain stellar core properties. Here, we showed that the phase offset of the gravity modes can help in the asymptotic regime to constrain the nature of the core (convective or radiative). However this phase offset is very sensitive to the inaccuracy of the asymptotic relation for the gravity modes. When the nature of the core is unambiguous, it can be used to delimit the validity of the asymptotic theory for the gravity modes. In giants and subgiants, no fusion reactions occur in the core and therefore it is necessarily radiative. 

In a subgiant such as KIC 11026764, the asymptotic relation for the gravity modes is a poor approximation because it is a first-order approximation and is valid only for high radial order. In RGBs, the mode order at maximum of amplitude is rather high, such that the asymptotic relation for the gravity modes is fairly accurate. KIC 12058433, being an early red giant, has a phase offset consistent with theoretical expectations.

\section{Conclusion} \label{sec:7}

Mixed modes are difficult to identify and fit because their frequencies do not follow the asymptotic relations for pressure or gravity modes. However, as shown in the present paper, they are not randomly distributed in frequency but rather follow a distinctive pattern. This pattern is a sum of several avoided crossings, each being due in subgiants to the coupling of pressure modes with a single gravity mode.

Methods and tools to identify mixed modes have been already proposed (e.g. \citealt{bedding2011b, Mosser2012a, Benomar2012b}). But among them the coupled harmonic oscillator model is the only one that relies on individual pressure and gravity mode frequencies (rather than global, average quantities such as $\Delta\nu$ or $\Delta\Pi_\ell$) and this allows to fit precisely the observed power spectrum.
However, with this model a Bayesian approach is necessary because it is very complex and has too many degrees of freedom. Without guidance the model could easily lead to non-physical results. 

In the present work, we used a set of priors that describe as best as possible our current knowledge, achieving  proper balance between prior and likelihood. The coupled oscillator model has been embedded in a MCMC fitting algorithm in order to both identify and fit the individual gravity, pressure and $\ell=1$ mixed modes of subgiant stars. 
Thanks to this, we were able to fit precisely all the properties (width, height and frequency) of the observed modes for four subgiants of different mass and evolutionary stage. The coupled oscillator model also permits us to infer the frequencies of the gravity-modes and therefore to measure accurately the period spacing of the gravity-modes.

\textbf{The coupled oscillator model allows us to infer where the $\ell=1$ pressure modes would be if pressure and gravity modes were uncoupled (referred to as the home ridge). These mode frequencies in return, permit us to measure the actual distance between the home ridge and the observed mixed modes, which is a proxy of the degree of mixing. We therefore measured the effect of the mixing on the main characteristics of the modes and found that the more gravity-like the modes, the lower the widths and amplitudes. This qualitatively confirms the results of \cite{Dupret2009}. 
We stress that the stochastic nature of the excitation of the solar-like oscillations complicates
measurements of mode heights. However, analysis of longer time-series should allow us better to constrain this parameter thanks to an increase of the frequency resolution.}

Finally, in addition to $\ell=1$ mixed modes, we also clearly identified an $\ell=2$ mixed mode in KIC 11026764 and showed that we are not in the asymptotic regime for gravity modes.

\textbf{\textit{Acknowledgments. Funding for the Stellar Astrophysics Centre is provided by The Danish National Research Foundation (Grant DNRF106). The research is supported by the ASTERISK project (ASTERoseismic Investigations with SONG and Kepler) funded by the European Research Council (Grant agreement no.:267864). We would like to thanks G\"ulnur Do{\u g}an and Travis Metcalfe for the fruitful discussions.}}

\clearpage

\bibliographystyle{apj}

\clearpage

\clearpage

\clearpage

\end{document}